\documentclass[aps,prd,showpacs,nofootinbib,superscriptaddress,amsmath,amssymb]{revtex4}

\usepackage{slashed}
\usepackage[dvips]{color}
\usepackage[dvips]{epsfig}
\usepackage{latexsym}
\usepackage{bm}
\usepackage{upgreek}
\usepackage{mathrsfs}
\usepackage{times}
\usepackage{amsthm}
\usepackage{amssymb}
\usepackage{epsfig}
\usepackage{graphicx}
\usepackage{amsmath}

\usepackage{color}

\definecolor{purple}{rgb}{0.8,0,0.6}
\newcommand{\revisionA}[1]{{#1}}

\newcommand{\revisionZ}[1]{{#1}}

\begin{document}

\title{Modified top quark condensation model with the extra heavy fermion, the $125$ GeV Pseudo - Goldstone boson, and the additional heavy scalar bosons}

\author{Z.V.Khaidukov}
\affiliation{Institute for Theoretical and Experimental Physics, B. Cheremushkinskaya 25, Moscow, 117259, Russia}

\author{M.A. Zubkov}
\email{zubkov@itep.ru}
\affiliation{LE STUDIUM, Loire Valley Institute for Advanced Studies,
Tours and Orleans, 45000 Orleans France}
\affiliation{Laboratoire de Mathematiques et de Physique
Theorique, Universite de Tours, 37200 Tours, France}
\affiliation{Institute for Theoretical and Experimental Physics, B. Cheremushkinskaya 25, Moscow, 117259, Russia}
\affiliation{National Research Nuclear University MEPhI (Moscow Engineering
Physics Institute), Kashirskoe highway 31, 115409 Moscow, Russia}

\begin{abstract}
We discuss the modified top quark condensation model proposed in \cite{VZ2015}. This construction was inspired by the top - seesaw scenario, in which the extra heavy fermion $\chi$ is added that may be paired with the top quark. Besides, this model incorporates the ideas of the Little Higgs scenario, in which the $125$ GeV scalar particle appears as a Pseudo - Goldstone boson.  This model admits (in addition to the $125$ GeV scalar boson $H$) the heavier scalar excitation $H^\prime$. We consider the region of parameters, where  its mass is $M_{H^\prime} \sim  1$  TeV, the width of $H^\prime$ is $\Gamma_{H^\prime} \sim 0.3 M_{H^\prime}$, while the mass of the heavy fermion is $m_\chi \sim 1$ TeV. We find that in this model the value of the cross - section $\sigma_{pp \to H^\prime + X \to \gamma+\gamma + X}$ for  $\sqrt{s}=13$ TeV is essentially smaller than the present experimental upper bound. Besides, we find, that  for the chosen values of parameters there should exist the CP - even scalar boson with mass $\approx 2 m_\chi$ and very small width. In addition, the model predicts the existence of the extra neutral CP even scalar boson and the charged scalar boson with  masses of the order of $1$ TeV.
\end{abstract}


\pacs{12.60.Fr  12.60.Rc  14.80.Bn 14.80.Ec\\
composite Higgs bosons, top - seesaw, LHC, proton - proton collisions}


\maketitle

\newcommand{\br}{{\bf r}}
\newcommand{\bu}{{\bf \delta}}
\newcommand{\bk}{{\bf k}}
\newcommand{\bq}{{\bf q}}
\def\({\left(}
\def\){\right)}
\def\[{\left[}
\def\]{\right]}

\newcommand{\barray}{\begin{eqnarray}}
\newcommand{\earray}{\end{eqnarray}}
\newcommand{\nn}{\nonumber \\}
\newcommand{\nl}{& \nonumber \\ &}
\newcommand{\bnl}{\right .  \nonumber \\  \left .}
\newcommand{\dbnl}{\right .\right . & \nonumber \\ & \left .\left .}

\newcommand{\beq}{\begin{equation}}
\newcommand{\eeq}{\end{equation}}
\newcommand{\ba}{\begin{array}}
\newcommand{\ea}{\end{array}}
\newcommand{\bea}{\begin{eqnarray}}
\newcommand{\eea}{\end{eqnarray} }
\newcommand{\be}{\begin{eqnarray}}
\newcommand{\ee}{\end{eqnarray} }
\newcommand{\bal}{\begin{align}}
\newcommand{\eal}{\end{align}}
\newcommand{\bi}{\begin{itemize}}
\newcommand{\ei}{\end{itemize}}
\newcommand{\ben}{\begin{enumerate}}
\newcommand{\een}{\end{enumerate}}
\newcommand{\bc}{\begin{center}}
\newcommand{\ec}{\end{center}}
\newcommand{\bt}{\begin{table}}
\newcommand{\et}{\end{table}}
\newcommand{\btb}{\begin{tabular}}
\newcommand{\etb}{\end{tabular}}
\newcommand{\bvec}{\left ( \ba{c}}
\newcommand{\evec}{\ea \right )}

\newcommand\e{{e}}
\newcommand\eurA{\eur{A}}
\newcommand\scrA{\mathscr{A}}

\newcommand\eurB{\eur{B}}
\newcommand\scrB{\mathscr{B}}

\newcommand\eurV{\eur{V}}
\newcommand\scrV{\mathscr{V}}
\newcommand\scrW{\mathscr{W}}

\newcommand\eurD{\eur{D}}
\newcommand\eurJ{\eur{J}}
\newcommand\eurL{\eur{L}}
\newcommand\eurW{\eur{W}}

\newcommand\eubD{\eub{D}}
\newcommand\eubJ{\eub{J}}
\newcommand\eubL{\eub{L}}
\newcommand\eubW{\eub{W}}

\newcommand\bmupalpha{\bm\upalpha}
\newcommand\bmupbeta{\bm\upbeta}
\newcommand\bmuppsi{\bm\uppsi}
\newcommand\bmupphi{\bm\upphi}
\newcommand\bmuprho{\bm\uprho}
\newcommand\bmupxi{\bm\upxi}

\newcommand\calJ{\mathcal{J}}
\newcommand\calL{\mathcal{L}}

\newcommand{\notyet}[1]{{}}

\newcommand{\sgn}{\mathop{\rm sgn}}
\newcommand{\tr}{\mathop{\rm Tr}}
\newcommand{\rk}{\mathop{\rm rk}}
\newcommand{\rank}{\mathop{\rm rank}}
\newcommand{\corank}{\mathop{\rm corank}}
\newcommand{\range}{\mathop{\rm Range\,}}
\newcommand{\supp}{\mathop{\rm supp}}
\newcommand{\p}{\partial}
\renewcommand{\P}{\grave{\partial}}
\newcommand{\yDelta}{\grave{\Delta}}
\newcommand{\yD}{\grave{D}}
\newcommand{\yeurD}{\grave{\eur{D}}}
\newcommand{\yeubD}{\grave{\eub{D}}}
\newcommand{\at}[1]{\vert\sb{\sb{#1}}}
\newcommand{\At}[1]{\biggr\vert\sb{\sb{#1}}}
\newcommand{\vect}[1]{{\bold #1}}
\def\R{\mathbb{R}}
\newcommand{\C}{\mathbb{C}}
\def\hvar{{\hbar}}
\newcommand{\N}{\mathbb{N}}\newcommand{\Z}{\mathbb{Z}}
\newcommand{\Abs}[1]{\left\vert#1\right\vert}
\newcommand{\abs}[1]{\vert #1 \vert}
\newcommand{\Norm}[1]{\left\Vert #1 \right\Vert}
\newcommand{\norm}[1]{\Vert #1 \Vert}
\newcommand{\Const}{{C{\hskip -1.5pt}onst}\,}
\newcommand{\sothat}{{\rm ;}\ }
\newcommand{\Range}{\mathop{\rm Range}}
\newcommand{\ftc}[1]{$\blacktriangleright\!\!\blacktriangleright$\footnote{AC: #1}}

\section{Introduction}

ATLAS and CMS announced in December of 2015 an excess of events in the $\gamma \gamma$ channel at the value of invariant mass $\approx 750$ GeV, which was interpreted in certain publications as  the indication of the existence of new particle \cite{ATLAS,CMS}. This announcement caused the revival of interest to the composite models of Higgs bosons. A lot of various models were discussed that pretend to describe the possible origin of this hypothetical particle (for the review see \cite{750REV}, where, in addition, the references are given to more than $300$ theoretical papers, which discuss the possible origin of this excess of events). The recent analysis of the experimental data strengthened the upper bound on the cross - section for the production of the hypothetical $750$ GeV scalar boson and its decay to two photons \cite{ATLASCMSnew,ATLASCMSnew_,ATLASCMSnew__,ATLASCMSnew___}. With this upper bound there is still no evidence of the deviation of experimental data from the Standard Model (SM). However, as before, those deviations are not excluded as long as the experimental upper bound on  $\sigma_{pp \to H^\prime + X \to \gamma\gamma +X}$ is not exceeded. Nevertheless, if we assume, that the second Higgs boson exists, its mass is not yet fixed.  According to the present experimental constraints \cite{ATLASCMSnew,ATLASCMSnew_,ATLASCMSnew__,ATLASCMSnew___} the admitted values of $\sigma_{pp \to H^\prime + X \to \gamma\gamma +X}$  are smaller than about $5$ fb for the resonance with $M_{H^\prime} \sim 750$ GeV (and $\Gamma_{H^\prime} \sim 0.05 M_{H^\prime}$), this upper bound is increased with the increase of the width and is decreased with the increase of $M_{H^\prime}$, so that it is around $1$ fb for the resonance with $M_{H^\prime} \sim 1600$  GeV (and $\Gamma_{H^\prime} \sim 0.05 M_{H^\prime}$).

Here we discuss the scenario, in which the hypothetical extra scalar bosons as well as the $125$ GeV scalar boson are composed of the top quark and the additional heavy fermion $\chi$. Such scenarios follow the analogy with superconductivity and superfluidity. Historically, the scenarios of such type \cite{Terazawa,Terazawa2} were proposed even earlier, than the more popular technicolor theory  \cite{technifermions,technifermions_}\footnote{The techniclolor  theory contains an additional set of fermions that interact with the Technicolor (TC) gauge bosons. This interaction is attractive and, similar to the BCS superconductor theory it leads to the formation of condensate. In order to provide the generation of the fermion masses in the TC the Extended Technicolor (ETC) interactions \cite{ETC,ETC_} are added that unfortunately do not pass the precision Electroweak tests. This occurs due to the flavor changing neutral currents and because of the contributions to the Electroweak polarization operators. The possible solution of this problem was discussed actively within the context of the so-called walking technicolor \cite{Walking,Walking_}.}.
The possibility that the Higgs boson is composed of the pair of top quark and anti top quark was discussed actively starting from \cite{top,top_} (see also  \cite{top2,top2_,top2__,top2___}).  Later the models were developed \cite{top3,top3_,top3__,top3___,top3____,top3_____,top3______}  that contain the elements of both top quark condensation scenario and technicolor. The idea that the Higgs boson appears as the Pseudo - Goldstone boson was proposed in \cite{PNG}. This idea was realized, in particular, in the  Little Higgs Models \cite{LittleHiggs,LittleHiggs_,LittleHiggs__} which became popular relatively recently. The composite Higgs bosons in the models, that contain in addition to the  top quark the extra heavy fermion $\chi$, were discussed in the framework of the top seesaw scenario \cite{topseesaw,topseesaw_}.

It is worth mentioning, that among the theoretical papers that appeared between December, 2015 and August, 2016  there are several ones, which consider both $125$ GeV boson and the hypothetical new heavier Higgs boson as composite due to the new strong interaction (see, for example, \cite{HF1,HF2}. More papers, however, were devoted to the description of the composite nature of the heavier ($750$ GeV) Higgs boson only \cite{F1,F2,F3,F4,F5,F6,F7,F8,F9,F10,F11,F12,F13,F14,F15,F16}.
In the present paper we concentrate on the scenario, which is based on the application of the model of \cite{VZ2015}. This model incorporates several mentioned above ideas that  existed earlier in the high - energy physics (top - quark condensation scenario, top - seesaw scenario, the appearance of the $125$ GeV Higgs boson as the Pseudo - Goldstone boson). Those three ingredients were also present in the two composite models proposed in \cite{dobrescu} and \cite{Yamawaki}. An extra ingredient, which was incorporated in \cite{VZ2015} is the analogy to the physics of $^3$He-B superfluid, where the pseudo - Goldstone boson appears with the Leggett frequency due to the soft breakdown of the basic symmetry by spin - orbit interactions\footnote{See also \cite{VolovikZubkov2014,VolovikZubkovHiggs,VolovikZubkovHiggs_}, where some other composite models of the Higgs bosons were proposed basing on an analogy to the models of condensed matter physics.}. According to this analogy the soft breakdown of the global $SU(3)_L$ symmetry is provided in \cite{VZ2015} by the four - fermion interaction without the use of the explicit mass term (this is in contrast to the models of \cite{dobrescu,Yamawaki}).  In the recent experimental paper \cite{V2016} devoted to the investigation of $^3$He this analogy was mentioned as well, and basing on this analogy the existence of new heavy composite Higgs bosons in the ultraviolet completion of the Standard Model was suggested. In the present paper we adopt the model of \cite{VZ2015} to the description of the possible common origin of the $125$ GeV scalar boson and the new hypothetical scalar particles. We fix for the definiteness the value of the mass of the second Higgs boson to the same value  $750$ GeV that was discussed in the mentioned above publications. In view of the results of \cite{ATLASCMSnew,ATLASCMSnew_} we do not consider this value as the preferred one, and take it only as an example. Besides, we represent here our results for the example choices of parameters with $M_{H^\prime} \approx 1200$ GeV, $M_{H^\prime} \approx 1600$ GeV, and $M_{H^\prime}\approx 2000$ GeV.  The values of  the mass of the heavy fermion $\chi$  were considered within the interval between $600$ GeV and $6000$ GeV.

It appears, that without the additional ingredients in the model of \cite{VZ2015} the value of  the cross - section for the process $pp\to H^\prime + X \to \gamma\gamma +X$  is much smaller than the upper bounds indicated by ATLAS and CMS \cite{750REV,ATLASCMSnew}. This  means, that this model is in accordance with the present experimental data on the $\gamma \gamma$ channel.

The paper is organized as follows.  In Section
\ref{seesaw} we briefly remind the construction of \cite{VZ2015}, in which the Pseudo - Goldstone boson composed  of top quark and the heavy fermion $\chi$ plays the role of the $125$ GeV Higgs boson. In Section \ref{SectPhenomenology} we present the effective lagrangian for the decays of the CP - even composite scalar bosons. In Section \ref{SectPhen} we discuss the phenomenology of the considered model: we consider the example choices of parameters and calculate the decay constants of the first and the second composite scalar bosons.  In Section \ref{sectconclusions} we end with the conclusions.

\section{The model under consideration}
\label{seesaw}


\subsection{$SU(3)$ symmetric lagrangian}

\revisionZ{In this section we shall describe briefly the model setup proposed in \cite{VZ2015}. This model follows the line of research based on the consideration of the top seesaw version discussed
in
\cite{dobrescu}. The model under consideration contains the SM fermions and the extra colored
 fermion $\chi$. It is supposed, that there is the hidden interaction between quarks and $\chi$ that may be taken into account effectively through the four - fermion terms. Neglecting the $SU(3)$ breaking terms and the gauge fields we come to the following form of the partition function:
 \begin{eqnarray}
 Z = \int D\bar{b}^\prime D b^\prime D\bar{t}^\prime D t^\prime D\bar{\chi}^\prime D\chi^\prime e^{i \int d^4 x L_f + i \int d^4 x L^{(4)}_I}
 \end{eqnarray}
 The two terms in the lagrangian $L_f$ and $L_I^{(4)}$ will be specified  below. Here  we denote by $b^{\prime},t^{\prime}$ and $\chi^{\prime}$ the Grassmann variables that compose Dirac spinors of the $b$ - quark, $t$ - quark, and the extra quark $\chi$. In this basis the quantum numbers of the left - handed $\chi^\prime_L$ and the right - handed $\chi^\prime_R$ including the hypercharge are equal to the quantum numbers of the right - handed top quark. (The gauge fields are not included into the above expression for the partition function. However, later we will discuss what will happen if they are included.) Correspondingly, $\bar{b}^{\prime},\bar{t}^{\prime}$ and $\bar{\chi}^{\prime}$ are the independent Grassmann variables representing the conjugate spinors. The kinetic fermion term in the lagrangian is:
 \begin{eqnarray}
 L_f = \bar{\psi}^\prime_L\Big(i\bar{\sigma}^\mu \partial_\mu \Big)\psi^\prime_L + \bar{\psi}^\prime_R\Big(i {\sigma}^\mu \partial_\mu \Big)\psi^\prime_R
 \end{eqnarray}
 Here $\sigma^\mu = (1, \sigma^1,\sigma^2,\sigma^3)$ and  $\bar{\sigma}^\mu = (1, -\sigma^1,-\sigma^2,-\sigma^3)$, where $\sigma^a$ for $a = 1,2,3$ are the Pauli matrices. We denoted the left  and right - handed components of spinors by
\begin{equation}
\psi^\prime_{L} = \left(\begin{array}{c}b^{\prime}_L\\t^{\prime}_{L} \\ {\chi}^{\prime}_L \end{array}
\right), \quad \psi^\prime_R =
 \left( \begin{array}{c}b^{\prime}_R\\t^{\prime}_R\\\chi^{\prime}_R  \end{array}\right)\label{btchi}
\end{equation}
In the following we also use notations $\psi^\prime_b = b^\prime, \psi^\prime_t = t^\prime, \psi^\prime_\chi = \chi^\prime$. The four - fermion interaction term has the following form:
\begin{eqnarray}
 L_I^{(4)} = \sum_{A = b,t,\chi}\sum_{B,\bar{B} = t,\chi}\Big(\bar{\psi}^\prime_{L,A}\psi^\prime_{R,B}\Big) \Big[\Omega^{-1}_{2\times 2}\Big]_{B\bar{B}} \Big(\bar{\psi}^\prime_{R,\bar{B}}\psi^\prime_{L,A}\Big)\label{LI04}
 \end{eqnarray}
We introduce here the $2\times 2$ matrix of coupling constants
\begin{equation}
\Omega_{2\times 2} = \left(\begin{array}{cc}\omega_{tt}& \omega_{t\chi}\\ \omega_{\chi t} & \omega_{\chi \chi}\end{array} \right)
\end{equation}
Here $\omega_{tt}$,  $\omega_{\chi\chi}$,
$\omega_{t\chi}=\omega_{\chi t}$ are the real - valued parameters of the dimension of mass squared. }

\revisionZ{The given model belongs to the class of the Nambu - Jona - Lasinio models that contain the four - fermion interactions. The standard way of dealing with these models is the introduction of the auxiliary   Hubbard - Stratonovich field. In our case this is the $3\times 3$ matrix $\Phi = (0, \Phi_t, \Phi_{\chi})$ composed of the two $3$ - component complex scalar fields $\Phi_t = (\Phi_{b_Lt_R},\Phi_{t_Lt_R},\Phi_{\chi_L t_R})^T$, $\Phi_\chi = (\Phi_{b_L\chi_R},\Phi_{t_L\chi_R},\Phi_{\chi_L \chi_R})^T$ \footnote{In the following we will also use alternative notations $\Phi_t = (\Phi_{bt},\Phi_{tt},\Phi_{\chi t})^T$, $\Phi_\chi = (\Phi_{b\chi},\Phi_{t\chi},\Phi_{\chi \chi})^T$ and $\Phi_t = (\Phi_{b_Lt},\Phi_{t_Lt},\Phi_{\chi_L t})^T$, $\Phi_\chi = (\Phi_{b_L\chi},\Phi_{t_L\chi},\Phi_{\chi_L \chi})^T$.}.
We rewrite the term $e^{i\int d^4 x L_I^{(4)}}$ entering the expression for the partition function as follows:
\begin{eqnarray}
e^{i\int d^4 x L_I^{(4)}} = {\rm const}\, \int D\Phi^+_t D\Phi_t D\Phi^+_\chi D\Phi_\chi e^{i \int d^4 x L_I}
\end{eqnarray}
where
\begin{equation}
L_I = -   {\rm Tr}\, \Phi  \Omega \Phi^+  - \Bigl[
\bar{\psi}^\prime_L
\Phi  \psi^\prime_R + (h.c.)\Bigr],\label{LI0}
\end{equation}
and
\begin{equation}
\Omega = \left(\begin{array}{cc}0&  0 \\ 0 & \Omega_{2\times 2}\end{array} \right) = \left(\begin{array}{ccc}0& 0 & 0 \\ 0 & \omega_{tt}& \omega_{t\chi}\\ 0 & \omega_{\chi t} & \omega_{\chi \chi}\end{array} \right)
\end{equation}
This interaction  term may be written explicitly as
\begin{eqnarray}
L_I &=& -\Big(\omega_t^{}\Phi_t^+ \Phi_t +
\omega_\chi^{} \Phi_{\chi}^+ \Phi_{\chi}\nonumber\\&& +
\omega_{t\chi} [\Phi_t^+
\Phi_{\chi}+\Phi_{\chi}^+\Phi_t]\Big) \nonumber\\&&- \Bigl[
\left(\begin{array}{ccc} \bar{b}^{\prime}_L & \bar{t}^{\prime}_L  & \bar{\chi}^{\prime}_L \end{array} \right)
\Phi_t t^{\prime}_R + \left(\begin{array}{ccc} \bar{b}^{\prime}_L & \bar{t}^{\prime}_L  & \bar{\chi}^{\prime}_L \end{array} \right)
\Phi_{\chi} {\chi}^{\prime}_R \nonumber\\&&+
(h.c.)\Bigr],\label{LI}
\end{eqnarray}
In the following the three components of $\psi$ will be denoted by  $b$, $t$, and $\chi$ in the basis, in which the mass matrix is diagonal (these are the true fields of $b$ - quark, top - quark and the new heavy quark $\chi$.}

As it was mentioned above, in the initial  basis we denote those components by $b^{\prime},t^{\prime}$ and $\chi^{\prime}$. In this basis the quantum numbers of $\chi^\prime_L$ and $\chi^\prime_R$ including the hypercharge (and the quantum numbers of $t^\prime_R$) are equal to the quantum numbers of the right - handed top quark. This is the doublet field $\left(\begin{array}{c}b^\prime_L\\t^\prime_L \end{array}\right)$, which is transformed under the $SU(2)_L$ SM gauge field. At the same time, the mass eigensatates are the mixtures of the states that are transformed in different way under the SM gauge group. Therefore, the gauge interactions of the SM break the $SU(3)_L$ symmetry of Eq. (\ref{btchi}), but as well as it was done in \cite{VZ2015}, we neglect this here.

The global symmetry of the given lagrangian is $SU(3)_L \otimes U(1)_L\otimes U(1)_{t, R}\otimes U(1)_{\chi, R}$. Here $SU(3)_L$ corresponds to the $SU(3)$ rotations of $\psi_L$, while  the $U(1)$ parts of the global symmetry of our lagrangian correspond to the transformations $\psi_L\rightarrow e^{i \alpha} \psi_L$, $\psi_{t,R} \rightarrow e^{i \beta}\psi_{t,R}$, and $\Phi_t \rightarrow e^{i (\alpha - \beta)} \Phi_t$ (and the similar transformation for $\chi$).

Using orthogonal rotation of $t_R$ and $\chi_R$ we may bring $\Omega$ to the
diagonal form, in which representation it is denoted by
\begin{equation}
\Omega^{(0)} = \left(\begin{array}{ccc}0 & 0 & 0 \\0 & \omega^{(0)}_t & 0 \\ 0 & 0& \omega^{(0)}_{\chi}
\end{array} \right)  \label{Omega0}
\end{equation}

\subsection{Soft breakdown of $SU(3)_L$}

Up to this point the description of the model followed that of  \cite{dobrescu}. (A similar construction has also been considered in \cite{Yamawaki}.). However, unlike \cite{dobrescu} in our approach the explicit mass terms $\sim \bar{\chi}_L t_R$ and $\sim \bar{\chi}_L\chi_R  $ are not added. Instead we restrict ourselves with the four - fermion interaction terms and do not consider the explicit mass term.
The following terms are added to the lagrangian (of the model written in the form with  the auxiliary field $\Phi$)
\begin{eqnarray}
L_G &=&  g^{(0)}_{\chi} |\Phi^3_{\chi}|^2 + g^{(0)}_{t}|\Phi^3_t|^2 + g^{(0)}_{t\chi} \Big( \bar{\Phi}^3_{\chi}\Phi^3_{t} + (h.c.) \Big) \nonumber\\ &=&  {\rm Tr}\, \Phi \, G^{(0)} \Phi^+ \Upsilon_3, \label{gtc3}
\end{eqnarray}
and
\begin{eqnarray}
L_B &=&  -b^{(0)}_{\chi} |{\rm Im}\Phi^3_{\chi}|^2 - b^{(0)}_{t}|{\rm Im}\Phi^3_t|^2 \nonumber\\&&- 2b^{(0)}_{t\chi}( {\rm Im}{\Phi}^3_{\chi})({\rm Im} \Phi^3_{t}) \nonumber\\ &=&  \frac{1}{4}{\rm Tr}\, (\Phi-\Phi^*) \, B^{(0)} (\Phi^T-\Phi^+) \Upsilon_3, \label{gtc4}
\end{eqnarray}
where
\begin{eqnarray}
G^{(0)}&=&\left(\begin{array}{ccc}0 & 0 & 0 \\0 & g^{(0)}_t & g^{(0)}_{t\chi} \\ 0 & g^{(0)}_{t\chi}& g^{(0)}_{\chi}
\end{array} \right) ,\quad B^{(0)}=\left(\begin{array}{ccc}0 & 0 & 0 \\0 & b^{(0)}_t & b^{(0)}_{t\chi} \\ 0 & b^{(0)}_{t\chi}& b^{(0)}_{\chi}
\end{array} \right) ,\nonumber\\  \Upsilon_3 & = & \left(\begin{array}{ccc}0 & 0 & 0 \\0 & 0 & 0 \\ 0 & 0& 1
\end{array} \right)\label{G3}
\end{eqnarray}
After the integration over $\Phi$ we arrive instead of Eq. (\ref{LI04}) at the four - fermion interaction of a more complicated form, which does not preserve the $SU(3)_L$ symmetry.

As it was mentioned above, we may bring $\Omega$ to the diagonal form via the orthogonal rotations of $\psi_R$. Further  the representation in this basis will be assumed. We also imply that the elements of matrices $\Omega$, $B$ and $G$ (that contain coupling constants) are real - valued.

Symmetry breaking pattern in the given model is as follows. Without the $SU(3)$ breaking terms we have the original global $SU(3)_L \otimes U(1)_L\otimes U(1)_{t, R}\otimes U(1)_{\chi, R}$ symmetry that is broken spontaneously down to $U(1)_{t}\otimes U(1)_{\chi}\otimes U(1)_b$. (Here $U(1)_{t}$, $U(1)_{\chi}$ act on the left and the right - handed components of $t$ and $\chi$  while $U(1)_b$ acts on the left - handed b - quark.) As a result among the $12$ (real - valued) components of ${\Phi}$ we have $8$ Goldstone bosons. In the notations of \cite{VZ2015} we have $4$ massless states that are composed of $b$ - quark: $H_t^{\pm}, H_{\chi}^{\pm}$, there are $3$ CP - odd massless states $A_t, \pi_{\chi}$ and $\frac{A_{\chi}m_{\chi} + \pi_t m_{t}}{\sqrt{m_t^2+m_{\chi}^2}}$, and there is one CP - even massless state $\frac{m_{\chi} h_{\chi} - m_{t} \varphi_t}{\sqrt{m_t^2 +
m_{\chi}^2}}$.

When the $SU(3)$ breaking modification of the model is turned on, the original symmetry is reduced to $SU(2)_L \otimes U(1)_L$. This symmetry is broken spontaneously down to $U(1)_b$. As a result we have $3$ exactly massless Goldstone bosons to be eaten by $W^{\pm}$ and $Z$, and $5$ Pseudo - Goldstone bosons.
Only one of those Pseudo - Goldstone bosons is the CP even neutral scalar. It is to be identified with the $125$ GeV scalar particle.

\subsection{Gap equation and  the basis of mass eigenstates}

The input parameters of the model are:
$\omega_t^{(0)}$, $\omega^{(0)}_\chi$, $g^{(0)}_t$, $g^{(0)}_{t\chi}$, $g^{(0)}_\chi$, $b^{(0)}_t$, $b^{(0)}_{t\chi}$, $b^{(0)}_\chi$, $\Lambda$, where $\Lambda$ is the ultraviolet cutoff to be used in the integrals over momenta in the effective model with the four - fermion interactions.

In \cite{VZ2015} the gap equation has been derived in the leading order in the $1/N_c$ expansion, which determines the condensate of the Hubbard - Stratonovich field $\Phi$, and the corresponding masses of the fermions. It appears, that the mass eigenstates $\psi$ are related to the initial fermions as follows
\begin{eqnarray}
&& \psi^\prime_L = \Theta \psi_L, \quad \psi^\prime_R = A \psi_R,\nonumber\\ &&
\Theta^T {\Phi} A = {\rm diag}(0,m_t,m_{\chi})+\tilde{\Phi}\label{rot}
\end{eqnarray}
where $\tilde{\Phi}$ has vanishing condensate and represents excitations above vacuum, while
\begin{eqnarray}
\Theta & = & {\rm exp}\, \Big(-i \theta \sigma^2 \Big), \quad A  = {\rm exp}\,
\Big(-i \alpha \sigma^2 \Big), \nonumber\\ \sigma_2 & = & \left(\begin{array}{ccc}1&0&0\\ 0& 0  & -i  \\
0 & i  & 0
\end{array}\right)\label{rot2}
\end{eqnarray}
As a result we come to the following form of gap equation with diagonal matrix $\hat{m}={\rm diag}(0,m_t,m_{\chi})$:
\begin{eqnarray}
&& A^T \Omega^{(0)} A   - A^T\, G^{(0)} \, A \, \hat{m} \Theta^T \, \Upsilon_3 \, \Theta \,\hat{m}^{-1}  \nonumber\\ &&= 2 N_c \,I_{\Lambda}(\hat{m})\label{gap4}
\end{eqnarray}
where $N_c=3$ while by $I_{\Lambda}(m)$ we denote the function given by the integral
\begin{eqnarray}
I_{\Lambda}(m) &=& \frac{i}{(2\pi)^4} \, \int d^4 l \, \frac{1}{l^2-
m^2}\label{Zeq}
\end{eqnarray}
We also introduce the integral
\begin{eqnarray}
I_{\Lambda}(m_1,m_2,p) &=& - \frac{i}{(2\pi)^4} \, \int d^4 l \, \frac{1}{(l^2-
m_1^2)[(p-l)^2-m_2^2]}\label{Zeq2}
\end{eqnarray}
Both Eqs. (\ref{Zeq}) and (\ref{Zeq2})  depend on the value of the effective ultraviolet cutoff $\Lambda$ of the theory with the four - fermion interactions. We do not specify here how it is incorporated into the theory in the particular regularization. The important point about the regularization is that the shift of the variable $l \to l + a$ may be made in the integral \cite{cvetic}. That means that this is not possible simply to cut the integrals at $|l| = \Lambda$ from the very beginning\footnote{The example of the regularization that allows the change of variables $l \to l + a$  is presented, for example, in \cite{cvetic}. Also the dimensional regularization may play such a role \cite{NJLdim,NJLdim_,NJLdim__}, but then $\Lambda$ is given by a certain combination of the dimensional parameter $\mu$ of the regularization and $\epsilon = 4-D$. The bare coupling constants entering the four - fermion interaction term in the dimensional regularization are  related to the bare parameters of the model in the other regularizations by a finite renormalization. Dzeta - regularization also admits the shift $l \to l+a$ (see, for example, \cite{Z2014,Z2014_,Z2013}).}. The possibility to apply the mentioned shift of the integration variable is used when we calculate the one - loop effective action for the composite scalar boson \cite{VZ2015}. As a result the quadratically divergent terms are cancelled in this effective action and one is left with the logarithmic divergent integrals. Those integrals are already simply cut at the (Euclidean) momentum $p^2  = \Lambda^2$.  This is a more or less standard procedure (see, for example, \cite{NJLQCD} and references therein). Within this procedure we have
\begin{eqnarray}
&& N_c I_\Lambda(m_a,m_b,p)  =  \\ &&\frac{N_c}{16 \pi^2} \int_0^1 dx \, {\rm log}\,\frac{\Lambda^2}{m_a^2 x + m_b^2 (1-x) - p^2 x (1-x)}\label{eq1}
\end{eqnarray}
and
\begin{eqnarray}
 N_c I_\Lambda(m_a,m_b,0) & = & \frac{N_cI_\Lambda(m_b) - N_c I_\Lambda(m_a)}{m_a^2-m_b^2}\label{eq2}
\end{eqnarray}

 The difference between the final results and the results obtained using the original regularization (that allows the shift $l  \to l+a$) disappears in the formal limit $\Lambda \to \infty$. Altogether this procedure of the calculation of effective action may be considered as the phenomenological low energy theory with a certain dimensional parameter $\Lambda$. In \cite{cvetic} it is demonstrated, that this procedure is consistent with the $1/N_c$ expansion in the NJL model if  $\Lambda$ is at most several times larger, than the dynamical quark mass  \footnote{In the example choice of parameters considered here we have $4 m_\chi  > \Lambda$}.

 Using the dimensional regularization one is able to obtain the relations Eq. (\ref{eq1}) and Eq. (\ref{eq2}) precisely. For this the effective cutoff is taken equal to
 \begin{equation}
\Lambda = \mu \, {\rm exp}\,\Big( \frac{1}{4-D} - \frac{\gamma_E}{2} + \frac{{\rm log}\,4\pi}{2}\Big)
\end{equation}
where $\mu$ is the dimensional parameter entering each integral over momenta through the combination $d^D p \mu^{4-D}$, while $D$ is the dimension of space - time; $\gamma_E$ is the Euler constant.
The similar situation takes place in zeta regularization. Moreover, in those two regularizations one is able to use formally the $1/N_c$ expansion in the complete model with finite value of $\Lambda$ and with any relation between $\Lambda$ and the dynamical masses of fermions (see, for example, \cite{Z2013,Z2014,VZ2015}).

The elements of   $\tilde{\Phi}$ are denoted by
\begin{eqnarray}
\tilde{\Phi} =
\left(\begin{array}{ccc} 0 & H_t^- &  H_{\chi}^- \\0 & \frac{1}{\sqrt{2}}( h_t +i
A_t) &  \frac{1}{\sqrt{2}}(h_{\chi} +i
A_{\chi})\\
 0 &  \frac{1}{\sqrt{2}}(\varphi_{t} +i \pi_{t}) &   \frac{1}{\sqrt{2}}(\varphi_{\chi} +i \pi_{\chi})
\end{array}\right)
\end{eqnarray}

By $g_{\chi}$, $g_t$, $g_{t\chi}=g_{\chi t}$ we denote the elements of matrix
\begin{equation}
G = A^T\, G^{(0)} \, A \label{GG0}
\end{equation}
 By $\omega_t$, $\omega_\chi$, $\omega_{t\chi}=\omega_{\chi t}$ we denote the elements of matrix
 \begin{equation}
 \Omega = A^T \Omega^{(0)} A\label{OO0}
 \end{equation}
The angles $\theta$, and $\alpha$, and the values of masses $m_t$, $m_\chi$ are to be determined through the solution of the following system of equations, which is accompanied by Eqs. (\ref{GG0}), (\ref{OO0}):
\begin{eqnarray}
 {\rm tg}\,2\theta & = & \frac{2\,g_{t\chi}}{g_\chi m_\chi/m_t-g_t m_t/m_\chi} \nonumber\\
 \omega_{t\chi} &=& \Big( g_t\, \frac{m_t}{m_{\chi}} \, {\rm sin}\,\theta +g_{t\chi} \, {\rm cos}\, \theta \Big)\, {\rm cos}\, \theta   \label{alpha}\\
{\omega}_{t}   &=& f_t + 2 N_c\,I_{\Lambda}(m_t)   \nonumber\\
 {\omega}_{\chi} &=& f_\chi + 2 N_c\,I_{\Lambda}(m_\chi),\label{gapeqs}
\end{eqnarray}
where
\begin{eqnarray}
f_t & = &{\rm sin} \, \theta\, \Big(g_t {\rm sin}\, \theta+g_{t\chi} \frac{m_{\chi}}{m_t}\, {\rm cos}\, \theta\Big) \nonumber\\ f_{\chi} &=& {\rm cos} \, \theta\, \Big(g_{t\chi}\frac{m_t}{m_{\chi}} {\rm sin}\, \theta+g_{\chi} \, {\rm cos}\, \theta\Big)\label{ff}
\end{eqnarray}
Unfortunately the final solution of this system of equations is so complicated, that we do not represent it here. Notice, that there exist the critical values of coupling constants that separate the region of parameters, for which the gap equation has the nonzero solution for $m_t$ and (or) $m_\chi$ from the region of parameters, where there is no such a solution.

By $b_{\chi}$, $b_t$, $b_{t\chi}=b_{\chi t}$ we denote the elements of matrix
\begin{equation}
B = A^T\, B^{(0)} \, A \label{BB0}
\end{equation}
These values may be calculated once the value of $\alpha$ is known.

\section{The Effective lagrangian for the decays of the CP - even scalar  bosons}

\label{SectPhenomenology}

\subsection{Higgs boson decay constants}

Typically the production cross - sections and the decays of the neutral Higgs boson $X$ are  described by the effective lagrangian of the following form:
\begin{widetext}
\begin{eqnarray}
\label{eq:12}
L_{eff}  &=  &
  c^H_W {2 m_W^2  \over \eta} H  \,   W_\mu^+ W_\mu^-  +     c^H_Z {m_Z^2 \over \eta} H\,  Z_\mu  Z_\mu
 + c^H_{g}  {\alpha_s \over 12 \pi \eta} H \, G_{\mu \nu}^a G_{\mu \nu}^a  +  c^H_{\gamma} { \alpha \over \pi \eta} H  \, A_{\mu \nu} A_{\mu \nu} \nonumber\\ &&- c^H_t \frac{m_t}{\eta}\bar{t}{t}\,H - c^H_\chi \frac{m_t}{\eta}\bar{\chi}{\chi} H - c^H_{\chi t} \frac{m_t}{\eta}(\bar{\chi}_L{t}_R + h.c.) H -  c^H_{t \chi} \frac{m_t}{\eta}(\bar{t}_L{\chi}_R + h.c.)H
\end{eqnarray}
\end{widetext}
Here $G_{\mu\nu}$ and $A_{\mu\nu}$ are the field strengths of gluon and photon fields. We do not consider here the masses of the fermions other than the top quark and $\chi$. Therefore, we omit in this lagrangian the terms responsible for the coresponding decays.
This effective lagrangian should be considered at the tree level only and describes the channels $H \rightarrow gg, \gamma \gamma, ZZ, WW, t \bar{t}, \chi \bar{\chi}, \chi \bar{t}, t \bar{\chi}$. The fermions and $W$ bosons have been integrated out in the terms corresponding to the decays $H\rightarrow \gamma \gamma, gg$, and their effects are included in the effective  couplings $c^H_g$ and $c^H_\gamma$. In the SM we have $c_t=c_Z=c_W = 1$, while $c_{g}   \simeq 1.03\,, c_{\gamma} \approx  -0.81$ (see \cite{status}).
In \cite{VZ2015} it was demonstrated, that the mentioned coupling constants for the CP even pseudo Goldstone boson coincide with that of the SM   $125$ GeV Higgs boson $H$ when the ratio $m_t/m_\chi$ is neglected.

In general case the constants for the decays to two photons and two gluons are related to the values of $c^H_t$ and $c^H_\chi$ as follows:
\begin{equation}
c^H_g=c^H_t {A_f\Big(M^2_{H}/(4 m_t^2)\Big)} + c^H_\chi{A_f\Big(M^2_{H}/(4 m_\chi^2)\Big)}\frac{m_t}{m_\chi}
\end{equation}
with
\begin{equation}
A_f(\tau) = \frac{3}{2\tau^2}\Big((\tau-1)f(\tau)+\tau\Big)
\end{equation}
and
\begin{eqnarray}
f(\tau) &=& {\rm arcsin}^2\sqrt{\tau}, \quad \tau < 1\nonumber\\
f(\tau) &=& -\frac{1}{4}\Big[ {\rm log}\frac{1+\sqrt{1-\tau^{-1}}}{1-\sqrt{1-\tau^{-1}}}-i \pi\Big]^2, \quad \tau > 1
\end{eqnarray}

$c^H_{\gamma}$ is given by
\begin{eqnarray}
c^H_{\gamma} &=& \frac{2}{9} c^H_{t} A_f\Big(\frac{M_{H}^2}{4 m_t^2}\Big) + \frac{2}{9} c^H_{\chi} A_f\Big(\frac{M_{H}^2}{4 m_\chi^2}\Big)\frac{m_t}{m_\chi} \nonumber\\&&- \frac{7}{8} c^H_{W} A_v\Big(\frac{M_{H}^2}{4 M_W^2}\Big)
\end{eqnarray}

\subsection{Calculation of neutral scalar boson masses}

Let us demonstrate how  to calculate the values of $c_t$, $c_\chi$. The decays of the field $\Phi$ to the pairs $t\bar{t}$ and $\chi \bar{\chi}$ are described by the lagrangians
\begin{equation}
L_{\Phi\rightarrow \bar{t}t} = - \Bigl[
\bar{t}_L
\Phi_{tt}  t_R + (h.c.)\Bigr]
\end{equation}
and
\begin{equation}
L_{\Phi\rightarrow \bar{\chi}\chi} = - \Bigl[
\bar{\chi}_L
\Phi_{\chi\chi}  \chi_R + (h.c.)\Bigr]
\end{equation}
The mass eigenstates of the fermions enter those expressions. Correspondingly the fields $\Phi_{tt}$ and $\Phi_{\chi\chi}$ are defined in this basis. Let us compose the four - component field out of the real parts of $\Phi_{ij}$:
\begin{equation}
\Phi = \left(\begin{array}{c}\Phi_{tt}\\ \Phi_{t\chi}\\ \Phi_{\chi t} \\ \Phi_{\chi\chi} \end{array}\right)\label{phi4}
\end{equation}
 The effective action for this field is given by
 \begin{equation}
 S_{\Phi} = \sum_{p}\Phi^T_p {\cal P}^{\prime}_{}(p^2) \Phi_p
 \end{equation}
where matrix ${\cal P}^{\prime}_{}(p^2)$ was calculated in \cite{VZ2015}. Let us represent it as follows:
\begin{equation}
{\cal P}^{\prime}_{}(p^2) = -p^2 \hat{Z}^2 + {\bf M}^2,
\end{equation}
where
\begin{widetext}
\begin{eqnarray}
{\bf M}^2 = \left(\begin{array}{cccc}
\begin{array}{c}4m_t^2\times\\\times N_c I_\Lambda(m_t,m_t,p)\\+f_t - g_t \lambda_t\end{array}& \omega_{t\chi} - g_{t\chi} \lambda_t&  -g_t \lambda_{t\chi} &-g_{t\chi}\lambda_{t\chi} \\
\omega_{t\chi} - g_{t\chi}\lambda_t&\begin{array}{c}( m_{t}^2+m_\chi^2)\times\\\times N_cI_\Lambda(m_t,m_\chi,p)\\ + (m_t^2-m_\chi^2)N_c I_\Lambda(m_\chi,m_t)\\+ f_{\chi}- g_{\chi} \lambda_t\end{array}&\begin{array}{c} 2 m_{t} m_{\chi}\times\\\times N_cI_\Lambda(m_t,m_\chi,p)\\ - g_{t\chi} \lambda_{t\chi}\end{array} & -g_{\chi}\lambda_{t\chi} \\
-g_t \lambda_{t\chi} & \begin{array}{c} 2 m_{t} m_{\chi}
\times\\\times N_cI_\Lambda(m_t,m_\chi,p)- g_{t\chi}\lambda_{t\chi}\end{array} &\begin{array}{c}( m_{t}^2+m_\chi^2)\times\\\times N_cI_\Lambda(m_t,m_\chi,p)\\ - (m_t^2-m_\chi^2)N_c I_\Lambda(m_\chi,m_t)\\+ f_t- g_t\lambda_{\chi}\end{array} &\omega_{t\chi}-g_{t\chi}\lambda_{\chi}
\\
-g_{t\chi}\lambda_{t\chi}&-g_{\chi}\lambda_{t\chi} & \omega_{t\chi}-g_{t\chi}\lambda_{\chi} & \begin{array}{c}4m_\chi^2\times\\\times N_c I_\Lambda(m_\chi,m_\chi,p)\\+f_{\chi}-g_{\chi} \lambda_{\chi}\end{array}
 \end{array}\right)
\label{mevenexact0}
\end{eqnarray}
\end{widetext}
where $$\lambda_{t}={\rm sin}^{2}\theta,\lambda_{\chi t}={\rm sin}\theta\,{\rm cos}\theta, \lambda_{\chi}={\rm cos}^2\theta$$
and
\begin{widetext}
\begin{eqnarray}
\hat{Z} = \left(\begin{array}{cccc}
\sqrt{ N_c I_\Lambda(m_t,m_t,p)}& 0& 0&0 \\
0& \sqrt{N_cI_\Lambda(m_t,m_\chi,p)}&0& 0\\
0&0 & \sqrt{N_cI_\Lambda(m_t,m_\chi,p) }&0
\\
0&0 &0&\sqrt{N_c I_\Lambda(m_\chi,m_\chi,p)}
 \end{array}\right)
\label{mevenexact3}
\end{eqnarray}
\end{widetext}

Next, we define
\begin{equation}
\hat{\Phi} = \hat{Z} \Phi
\end{equation}
 and $\hat{\bf M}^2 = \hat{Z}^{-1}{\bf M}^2\hat{Z}^{-1}$. As a result the effective action receives the form
\begin{equation}
 \hat{S}_{\Phi} = i \sum_p {\rm log}\,\hat{Z}(p) +\sum_{p}\hat{\Phi}^T_p (-p^2 + \hat{\bf M}^2(p)) \hat{\Phi}_p
 \end{equation}
 Matrix $\hat{\bf M}^2$ may be diagonalized using transformations:
$\hat{\bf m}^2 = O^T \hat{\bf M}^2 O = {\rm diag}\,(m^2_{H_1}(p),m^2_{H_2}(p),m^2_{H_3}(p),m^2_{H_4}(p))$ is diagonal. For Euclidean momenta $p^2<0$ matrices $O$ are orthogonal. For $p^2>0$ the elements of this matrix are complex - valued, but it obeys $O^T O =1$ anyway.  The masses of the scalar excitations may be found through the equations
\begin{equation}
p^2 = m^2_{H_i}(p)
\end{equation}
In the present section we limit ourselves with the consideration of neutral scalar bosons. The procedure for the calculation of the charged scalar boson masses and pseudo - scalar boson masses was developed in \cite{VZ2015}.

\subsection{Calculation of $c^H_t$ and $c^H_\chi$}

The columns of the matrix $O$ form the eigenvectors of matrix $\hat{\bf M}^2$:
\begin{eqnarray}
O = \left(\begin{array}{cccc}
{\bf o}_{H_1}(p) & {\bf o}_{H_2}(p)& {\bf o}_{H_3}(p)& {\bf o}_{H_4}(p)
 \end{array}\right)
\label{o}
\end{eqnarray}
The eigenvectors ${\bf o}_{H_i}(p) = (o_{H_i}^1,o_{H_i}^2,o_{H_i}^3,o_{H_i}^4)^T$ are normalized in such a way, that $${\bf o}^T_{H_i} {\bf o}_{H_j} = \sum_k { o}^k_{H_i} { o}^k_{H_j} = \delta_{ij}$$
For the space - like momenta with $p^2<0$ and for the time - like momenta below the threshold the values of ${o}^k_{H_i}$ are real, while those values may become complex above the threshold, when $p^2 > 4 m_t^2$.
The effective action now receives the form
\begin{equation}
 \hat{S}_{\Phi} = i \sum_p {\rm log}\,\hat{Z}(p) +\frac{1}{2}\sum_{p}{H}^T_p (-p^2 + \hat{\bf m}^2(p)) {H}_p
 \end{equation}
where
\begin{equation}
H = \sqrt{2}O^T \hat{\Phi} = \sqrt{2}O^T \hat{Z} \Phi
\end{equation}
The inverse relation is
\begin{equation}
\Phi = \frac{1}{\sqrt{2}} \hat{Z}^{-1} O H
\end{equation}
This results in
\begin{equation}
L_{\Phi\rightarrow \bar{t}t} = - \sum_i\frac{1}{\sqrt{2}Z_{ttH_i}}\Bigl[
\bar{t}_L
o^1_i H_i  t_R + (h.c.)\Bigr]
\end{equation}
and
\begin{equation}
L_{\Phi\rightarrow \bar{\chi}\chi} = - \sum_i\frac{1}{\sqrt{2}Z_{\chi\chi H_i}} \Bigl[
\bar{\chi}_L
o^4_i H_i  \chi_R + (h.c.)\Bigr]
\end{equation}
Here the sum is over all existing composite scalar fields $H_i$, which are called also in the text $H$, $H^\prime$, $H^{\prime \prime}$, $H^{\prime \prime \prime}$. At the same time $Z_{ttH}=\sqrt{N_c I_\Lambda(m_t,m_t,M_H)}$ and  $Z_{\chi\chi H}=\sqrt{N_c I_\Lambda(m_\chi,m_\chi,M_H)}$.
This gives
\begin{equation}
c^H_t = \frac{o^1_H}{Z_{t t H}} , \quad c^H_\chi =  \frac{o^4_H}{Z_{\chi\chi H}}
\end{equation}
In the similar way one may calculate the coupling constants
\begin{equation}
c^H_{t\chi} = \frac{o^2_H}{Z_{t \chi H}} , \quad c^H_{\chi t} =  \frac{o^3_H}{Z_{\chi t H}}
\end{equation}
where $Z_{t\chi H}=Z_{\chi t H}=\sqrt{N_c I_\Lambda(m_t,m_\chi,M_H)}$.

\subsection{Calculation of $c_W^H$ and $c_Z^H$}

In order to calculate $c_W^H$ let us recall, that the components of the scalar field (in the basis of the mass eigenstates of fermions) should be expressed through its components in the basis of the states, which experience weak interactions. The mass eigenstates $\chi_L$ and $t_L$ are composed of the original $\chi^{\prime}_L$ and $t^{\prime}_L$:
\begin{eqnarray}
\chi_L &=& - {\rm sin}\, \theta \, t^{\prime}_L +  {\rm cos}\, \theta \, \chi^{\prime}_L\nonumber\\
t_L &=&  {\rm }\,{\rm cos}\, \theta \, t^{\prime}_L +  {\rm sin}\, \theta \, \chi^{\prime}_L
\end{eqnarray}
This is the field $\left(\begin{array}{c}b^\prime_L\\t^{\prime}_L\end{array}\right)$, which carries the quantum numbers of the SM $SU(2)_L$ left - handed doublets. At the same time $t^\prime_R$, $\chi^\prime_L$, $\chi^\prime_R$ carry the quantum numbers of the right - handed top  quark.
Correspondingly, we represent
\begin{eqnarray}
\Phi_{\chi t} &=& - {\rm sin}\, \theta \, \Phi_{t^\prime_L t} +  {\rm cos}\, \theta \, \Phi_{\chi^{\prime}_L t}\nonumber\\
\Phi_{\chi \chi} &=& - {\rm sin}\, \theta \, \Phi_{t^\prime_L \chi} +  {\rm cos}\, \theta \, \Phi_{\chi^{\prime}_L \chi}\nonumber\\
\Phi_{t t}&=&  {\rm }\,{\rm cos}\, \theta \, \Phi_{t^{\prime}_L t} +  {\rm sin}\, \theta \, \Phi_{\chi^{\prime}_L t}
\nonumber\\
\Phi_{t \chi}&=&  {\rm }\,{\rm cos}\, \theta \, \Phi_{t^{\prime}_L \chi} +  {\rm sin}\, \theta \, \Phi_{\chi^{\prime}_L \chi}
\end{eqnarray}
 The four - vector of Eq. (\ref{phi4}) is expressed through the corresponding vector $\tilde{\Phi} =( \Phi^\prime_{t^\prime_L t },  \Phi^\prime_{t^\prime_L \chi},  \Phi^\prime_{\chi^\prime_L t}, \Phi^\prime_{\chi^\prime_L \chi})^T$ as
 \begin{equation}
 \Phi = \Theta \tilde{\Phi}, \quad \Theta = \left(\begin{array}{cccc}
 {\rm cos} \theta& 0& {\rm sin} \theta&0 \\
0& {\rm cos} \theta&0& {\rm sin} \theta\\
-{\rm sin} \theta&0 & {\rm cos} \theta&0
\\
0&-{\rm sin} \theta &0& {\rm cos} \theta
 \end{array}\right)
\end{equation}

\revisionZ{In order to calculate the W and Z boson masses in our model in the leading order of $1/N_c$ expansion this is necessary to calculate the terms in effective action in the presence of external $SU(2)\otimes U(1)$ gauge field up to the second order in the gauge field. To calculate those terms one should expand the fermion Determinant in the presence of external gauge field in powers of this field. This procedure leads to rather complicated nonlinear equations to be solved (for the description of the method see, for example, \cite{top2}). }

\revisionZ{Instead we use for the  estimate of $M_W$ and $M_Z$ the simplified method based on the construction of the low energy effective lagrangian. Namely, we approximate our model by the effective theory with the action
 \begin{equation}
 S_{\Phi} = \int d^4 x \Big( \Phi^+ \hat{Z}_0^2 \Box  \Phi - U(\Phi )\Big) \label{SEFF}
 \end{equation}
 Here $U(\Phi)$ is the effective potential, which has its minimum at $\Phi = (m_t,0,0,m_\chi)^T$ and provides the appearance of correct masses of these excitations. At the same time $\hat{Z}_0^2$ is defined as the value of matrix $\hat{Z}^2(p)$ at $p=0$. Eq. (\ref{SEFF}) gives the proper estimate of the effective low energy theory of the scalar field, which incorporates both its condensation and the masses of excitations above the condensate. The interaction with gauge field $A_\mu$ is then introduced gauging the derivative $\partial_\mu \rightarrow \partial_\mu - i A_\mu$.}

The kinetic part of effective action that gives masses of the gauge bosons may be represented as follows
 \begin{equation}
 S^{p^2}_{\Phi} = \sum_{p}\Phi^T_p \hat{Z}_0^2 p^2
  \Phi_p = \sum_p \tilde{\Phi}^T_p \Theta^T\hat{Z}_0^2p^2\Theta \tilde{\Phi} \label{SEFF1}
 \end{equation}
 When acting on $ \Phi^\prime_{t^\prime_L t },  \Phi^\prime_{t^\prime_L \chi}$ the operator $\hat{p}^2$ is to be substituted by the gauge field squared $A^2 = \frac{1}{4}(2g_W^2 W^+_\mu W^\mu + g^2_Z Z_\mu Z^\mu)$. Using relations $M_Z = g_Z \eta/2$ and $M_W = g_W \eta/2$ we come to the part of effective action, which contains the terms responsible for the interaction of the scalar fields with $W$ and $Z$ bosons.
 \begin{equation}
 S^{A^2}_{\Phi} = \sum_p \tilde{\Phi}^T_p \Pi_t \Theta^T\hat{Z}_0^2\Theta \Pi_t\tilde{\Phi}\Big(2\frac{M_W^2}{\eta^2} W W^+ + \frac{M_Z^2}{\eta^2} Z^2\Big)
 \end{equation}
where  \begin{equation}
 \Pi_t= \left(\begin{array}{cccc}
 1& 0& 0&0 \\
0& 1&0& 0\\
0&0 & 0&0
\\
0&0 &0& 0
 \end{array}\right)
\end{equation}
Next, we express this action in terms of $\Phi$:
 \begin{equation}
 S^{A^2}_{\Phi} = \sum_p {\Phi}^T_p {\cal W}{\Phi}\Big(2\frac{M_W^2}{\eta^2} W W^+ + \frac{M_Z^2}{\eta^2} Z^2\Big)\label{SA}
 \end{equation}
 where 
 ${\cal W} =   \Theta \Pi_t \Theta^T\hat{Z}_0^2\Theta \Pi_t \Theta^T$.
 Therefore,
 \begin{widetext}
 \begin{equation}
{\cal W} = \left(\begin{array}{cccc}
\begin{array}{cc}{\rm cos}^4 \theta  Z_{tt0}\\+{\rm cos}^2 \theta  {\rm sin}^2 \theta  Z_{t\chi 0}\end{array}& 0& \begin{array}{cc}-{\rm cos}^3 \theta  Z_{tt0} {\rm sin} \theta \\-{\rm cos} \theta  {\rm sin}^3 \theta  Z_{t\chi 0}\end{array}&0 \\
0&\begin{array}{cc}{\rm cos}^4 \theta  Z_{t\chi 0}\\+{\rm cos}^2 \theta {\rm sin}^2 \theta  Z_{\chi \chi 0}\end{array}&0& \begin{array}{cc}-{\rm cos}^3 \theta  Z_{t\chi 0} {\rm sin} \theta\\ -{\rm cos} \theta {\rm sin}^3 \theta  Z_{\chi \chi 0}\end{array}\\
\begin{array}{cc}-{\rm cos}^3 \theta  Z_{tt0} {\rm sin} \theta\\ -{\rm cos} \theta {\rm sin}^3 \theta  Z_{t\chi 0}\end{array}&0 &\begin{array}{cc} {\rm cos}^2 \theta  {\rm sin}^2 \theta  Z_{tt0}\\+{\rm sin}^4 \theta  Z_{t\chi 0}\end{array}&0\\
0&\begin{array}{cc}-{\rm cos}^3 \theta  Z_{t\chi 0} {\rm sin} \theta\\ -{\rm cos} \theta  {\rm sin}^3 \theta  Z_{\chi \chi 0} \end{array}&0& \begin{array}{cc}{\rm cos}^2 \theta  {\rm sin}^2 \theta  Z_{t\chi 0}\\+{\rm sin}^4 \theta  Z_{\chi \chi 0}\end{array} \end{array}\right) \nonumber
 \end{equation}
 \end{widetext}

Since $\langle \Phi \rangle = (m_t,0,0,m_\chi)^T$ we obtain the following expression (that relates parameters of the model with $\eta$) from the requirement, that $W$ and $Z$ bosons acquire the observable masses:
\begin{eqnarray}
\eta^2 &=& 2m_t^2 {\rm cos}^2 \theta (Z_{tt 0}^2 {\rm cos}^2 \theta + Z_{t\chi 0}^2 {\rm sin}^2 \theta) \nonumber\\&&+ 2 m^2_{\chi}\, {\rm sin}^2 \theta\, ( Z_{\chi\chi 0}^2 {\rm sin}^2\theta + Z_{t\chi 0}^2 {\rm cos}^2 \theta )   \nonumber\\ &\approx & 2 Z_{tt 0}^2 m_t^2 \Big(1 + \frac{g^2_{t\chi}}{g^2_{\chi}}\frac{Z_{t\chi 0}^2}{Z_{tt 0}^2}\Big) + O(m^2_t/m^2_\chi)\label{eta}
\end{eqnarray}

\revisionZ{In order to evaluate the accuracy of this estimate of $\eta$ we may compare our result with that of extracted from another possible version of the effective low energy theory for $\Phi$. In this version instead of the first term in Eq. (\ref{SEFF}) that is Eq. (\ref{SEFF1}) the kinetic term has the form (in which it is taken into account that mass operator depends logarithmically on $p$):
\begin{equation}
 S^{p^2}_{\Phi} = \sum_{p}\Phi^T_p \Big(\hat{Z}_0^2 - \frac{d}{d p^2}{\bf M}^2(p)\Big|_{p=0}\Big)p^2 \Phi_p \label{SEFF2}
 \end{equation}
 This gives the estimate of the accuracy of the obtained value of $\eta$ within about $10$ per cent for the example choices of parameters considered in the next section. Notice, that while dealing with both Eq. (\ref{SEFF1}) and Eq. (\ref{SEFF2}) we neglect the logarithmic dependence in Eq. (\ref{SEFF}) of $U(\Phi)$ on $\Box$. This dependence appears on the same grounds as the dependence of ${\bf M}$ on $p$. The corresponding $p$ -  depending terms are of the same order of magnitude. That is why the difference between the values of $\eta$ extracted from Eq. (\ref{SEFF}) and Eq. (\ref{SEFF2}) gives the accuracy of our evaluation of $\eta$.   }

The terms in Eq. (\ref{SA}) with the first power of the physical scalar field $H$ are given by
\begin{eqnarray}
 S^{H,A^2}_{\Phi}& =& H\Big(2\frac{M_W^2}{\eta} W W^+c^H_W + \frac{M_Z^2}{\eta} Z^2 c^H_Z\Big)\\
c^H_W &=& c^H_Z= \Big(\frac{{\rm cos}^4 \theta Z_{ttH}+{\rm cos}^2 \theta {\rm sin}^2 \theta Z_{t\chi H}}{\sqrt{Z_{ttH}}} o^1_H\nonumber\\&&+\frac{-{\rm cos}^3 \theta Z_{ttH} {\rm sin} \theta-{\rm cos} \theta {\rm sin}^3 \theta Z_{t\chi H} }{\sqrt{Z_{t\chi H}}} o^3_H\Big)\nonumber\\&&+\frac{m_\chi }{m_t} \Big(\frac{-{\rm cos}^3 \theta Z_{t\chi H} {\rm sin} \theta-{\rm cos} \theta {\rm sin}^3 \theta Z_{\chi \chi H} }{\sqrt{Z_{t\chi H}}} o^2_H\nonumber\\&&+\frac{{\rm cos}^2 \theta {\rm sin}^2 \theta Z_{t\chi H}+{\rm sin}^4 \theta Z_{\chi \chi H} }{ \sqrt{Z_{\chi \chi H}}} o^4_H\Big)
\end{eqnarray}

\section{Phenomenology}
\label{SectPhen}

\subsection{Partial widths}

In order to evaluate the branching ratio $Br({H^\prime} \to \gamma \gamma)$ we estimate the partial widths for the decays of ${H^\prime}$. For the ${H^\prime} \to gg$ decay we have:
\begin{equation}
 \Gamma_{{H^\prime} \to gg}
 = \left( \frac{\alpha_s}{6 \pi} \right)^2
 \cdot  \frac{M_{{H^\prime}}^3 }{4 \pi m_t^2 } \left| {c^{H^\prime}_g}\right|^2
 \approx 25 \text{ MeV} \times \left| {c^{H^\prime}_g} \right|^2\times \Big(\frac{M_{{H^\prime}} }{750\,{\rm GeV} }\Big)^3 ,
\end{equation}
For the process ${H^\prime} \to \gamma \gamma$ we have
\begin{equation}
 \Gamma_{{H^\prime} \to \gamma \gamma}
 = \left( \frac{\alpha}{\pi} \right)^2
   \frac{M_{H}^3}{8 \pi m_t^2}  \lvert c_\gamma^H \rvert^2
   \approx 3.6\, \text{ MeV}\times |c^{H^\prime}_\gamma|^2\times \Big(\frac{M_{{H^\prime}} }{750\,{\rm GeV} }\Big)^3
 \label{S->2gamma-width}
\end{equation}
Here the fine structure constant is given by $\alpha(M_{{H^\prime}}^2) = 1/125$.
The tree level estimate for the process ${H^\prime} \rightarrow t\bar{t}$ is
\begin{eqnarray}
 \Gamma_{{H^\prime} \to t \bar{t}}
 &=&  \frac{3}{16 \pi}\, M_{{H^\prime}}\, |c^{H^\prime}_t|^2 \Big(1-\frac{4 m_t^2}{M_{{H^\prime}}^2}\Big)^{3/2}
 \label{S->2t-width}
\end{eqnarray}
In particular, for $M_{H^\prime}= 750 $ GeV this estimate gives $ \Gamma_{{H^\prime} \to t \bar{t}}
   \approx  31\, \text{ GeV} \times |c^{H^\prime}_t|^2\times $.

The decay of ${H^\prime}$ to $b\bar{b}$ goes through the exchange by the virtual charged scalar whose mass was denoted in \cite{VZ2015} by $M^{ (2)}_{H^{\pm}_t, H^{\pm}_{\chi}}$. The transition between $b$ and $t$ is accompanied by the ejection of the corresponding scalar particle. This is not completely clear are we able or not to apply the perturbation theory for the calculation of the interaction vertex $b\bar{b} {H^\prime}$. Therefore, we represent here
 \begin{eqnarray}
 \Gamma_{{H^\prime} \to b \bar{b}}
 &=&  \frac{3 \zeta_{bb{t}{t}}}{16 \pi}\, M_{{H^\prime}}\, |c^{H^\prime}_t|^2 \Big(1-\frac{4 m_b^2}{M_{{H^\prime}}^2}\Big)^{3/2}\nonumber\\&
   \approx & 45\, \text{ GeV} \times \zeta_{bb{t}{t}}|c^{H^\prime}_t|^2\times \Big(\frac{M_{{H^\prime}} }{750\,{\rm GeV} }\Big)
 \label{S->2b-width}
\end{eqnarray}
where $\zeta_{bb{t}{t}}$ is the effective coupling constant that encodes the diagram with the four external lines corresponding to $b$, $\bar{b}$, ${t}$, $\bar{t}$, and the integral over momentum that corresponds to joining of the lines with $t$ and $\bar{t}$ at the point, where $H^\prime$ is created. In principle, we may add to our phenomenological model the four - fermion terms to the action, which give rise directly to this diagram. Therefore, $\zeta_{bb{t}{t}}$ here  may actually be considered as the phenomenological parameter. This parameter may be estimated using the value of the decay constant of the $125$ GeV Higgs boson $H$ into the pair $\bar{b} b$: $\Gamma_{H \to b \bar{b}} \approx \frac{3 \zeta_{bb{t}{t}}}{16 \pi}\, M_{{H}}\, \Big(1-\frac{4 m_b^2}{M_{{H}}^2}\Big)^{3/2}$ .  We know, that this value is to be close to the one predicted by the SM. This gives  $$\zeta_{bb{t}{t}}\approx \frac{m_b^2}{m_t^2}$$
Taking into account that $\zeta_{bbtt} \ll 1$ we neglect the contribution of this decay to the total width.

Thus our model predicts the total width
 $$ \Gamma^{H^\prime}_{tot} \approx \Gamma_{{H^\prime} \to t \bar{t}}$$
 Notice, however, that the above tree level estimate for $\Gamma_{{H^\prime} \to t \bar{t}}$ differs from the estimates of  $ \Gamma^{H^\prime}_{tot}$ given in Tables \ref{table1},   \ref{table2},  \ref{table4},  \ref{table5}, \ref{table6},  \ref{table7}. The reason is that in those Tables the  estimate for the total decay width is given through the imaginary part of the masses of $H^\prime$, while those masses appear as the  zeros of  function ${\cal P}^{\prime}_{}(p^2)$. This function contains the resummed diagrams. Therefore, the estimate of the decay width presented in Tables  \ref{table1},   \ref{table2},  \ref{table4},  \ref{table5}, \ref{table6},  \ref{table7} is more precise than that of the tree - level estimate of Eq. (\ref{S->2t-width}).

\subsection{Cross  - section for the process $pp \rightarrow H^\prime+X $ that goes through the gluon fusion and the annihilation of the botton quarks}

According to \cite{TEVrev2} all partons existing in proton may be able to annihilate with the creation of the neutral scalar bosons. The corresponding cross - section is given by
\begin{equation}
 \sigma_{pp \to {H^\prime}}=\sum_{\cal P} \sigma^{({\cal P}\bar{\cal P})}_{pp \to {H^\prime}}=\sum_{\cal P}\frac{1}{M_H s}[C_{{\cal P}\bar{\cal P}}\Gamma_{{H^\prime} \rightarrow {\cal P}\bar{\cal P}}]K_{{\cal P}\bar{\cal P}}\label{sigmaK}
\end{equation}
where ${\cal P}={u,d,s,c,b,t,g,\gamma}$ denotes the partons, $\Gamma_{{H^\prime} \rightarrow {\cal P}\bar{\cal P}}$ is the width for the decay of ${H^\prime}$ to the given partons, while $\sqrt{s}=13$ TeV. The values of the partonic integrals $C_{\cal PP}$ are given in Eq. (4) of \cite{TEVrev2}:
\begin{eqnarray}
 C_{gg} &=& \frac{\pi^2}{8}
   \int\limits_{M_{H^\prime}^2/s}^{1}\frac{dx}{x}
    g(x)g(M_{H^\prime}^2/(sx))\nonumber\\
    C_{\gamma\gamma} &=& 8\pi^2
   \int\limits_{M_{H^\prime}^2/s}^{1}\frac{dx}{x}
    \gamma(x)\gamma(M_{H^\prime}^2/(sx))\nonumber\\
    C_{q\bar{q}} &=& \frac{4\pi^2}{9}
   \int\limits_{M_{H^\prime}^2/(sx)}^{1}\frac{dx}{x}
   \Big (\bar{q}(x)q(M_{H^\prime}^2/(sx))\nonumber\\&&+q(x)\bar{q}(M_{H^\prime}^2/(sx))\Big)
 \label{gg-luminosity2}
\end{eqnarray}
Here $g(x)$, $\gamma(x)$, $q(x)$, and $\bar{q}(x)$ are the parton distributions of gluons, photons, quarks, and anti-quarks correspondingly. \revisionA{We calculate parton distributions using the MSTW2008NLO  package \cite{partons}.}
 For our (rough) estimate we take the factor $K_{{\cal P}\bar{\cal P}}\approx 1.5$ (that accounts for the gluon corrections) following \cite{TEVrev2} for the gluon - gluon channel and $K_{{\cal P}\bar{\cal P}}\approx 1.2$ for the quark - anti quark channel .

In our model there exist the contributions from the gluon - gluon fusion, the annihilation of $b$ and $\bar{b}$ and the photon - photon fusion. The latter is suppressed by the smallness of the fine structure coupling constant as well as by the small value of $C_{\gamma \gamma}$. \revisionA{For the gluon fusion at $\sqrt{s} = 13$ TeV we take the value $C_{gg} = 2137$  \cite{TEVrev2} at $M_{H^\prime}=750$ GeV, we obtain using the MSTW2008NLO package the values $C_{gg} = 174$  at $M_{H^\prime}=1200$ GeV, $C_{gg} = 31$ at $M_{H^\prime}=1600$ GeV, and $C_{gg} = 7$ at $M_{H^\prime}=2000$ GeV. For example, at $M_{H^\prime} = 750$ GeV this gives}
\begin{equation}
 \Gamma_{{H^\prime} \to gg}
\approx 25.4 \text{ MeV} \times \left| {c^{H^\prime}_g} \right|^2 ,
\end{equation}
We assume  $\alpha_s(M_{H^\prime}) =
0.090$ and obtain for $M_{H^\prime} \sim 750$ GeV
\begin{equation}
 \sigma^{(gg)}_{pp \to {H^\prime} X} \approx 250\,fb \times \left| c^{H^\prime}_g \right|^2
 \label{productiongg}
\end{equation}

\revisionA{For the annihilation of the b - quarks we have
  $C_{bb}\sim 15$  at $M_{H^\prime} = 750$ GeV, $C_{bb}\sim 1.2$  at $M_{H^\prime} = 1200$ GeV, $C_{bb}\sim 0.2$  at $M_{H^\prime} = 1600$ GeV, and $C_{bb}\sim 0.04$  at $M_{H^\prime} = 2000$ GeV. (Again, we use the MSTW2008NLO package.)}  We also use that
\begin{equation}
\frac{1}{{\rm TeV}^2} \approx 0.389\,10^{6}\,{\rm fb}
\end{equation}
and\revisionA{, for example, at $M_H = 750$ GeV we}  obtain
\begin{equation}
\sigma^{({bb})}_{pp \to {H^\prime}} \approx 2.5 \times 10^3\,  \zeta_{bb{t}{t}}|c^{H^\prime}_t|^2\, {\rm fb}\label{productionbb}
\end{equation}

\subsection{Cross - section for the process $pp\to H^\prime + X \to \gamma \gamma + X$ }

In the model of \cite{VZ2015}, which was considered without any extensions, the value of the cross section $\sigma_{pp \to H^\prime +X \to \gamma \gamma + X }$ is given by the product of the branching ratio
\begin{equation}
 Br({H^\prime} \to \gamma \gamma)
 \approx \frac{3.6 \,{\rm MeV} \, |c_\gamma^{H^\prime}|^2}{31\,{\rm GeV}\,|c_t^{H^\prime}|^2 }
  \label{S->2gamma-branching}
\end{equation}
and a production cross - section given by the sum of the two terms
$\sigma^{({bb})}_{pp \to {H^\prime}} $ and $\sigma^{({gg})}_{pp \to {H^\prime}} $.
  As a result \revisionA{for $M_{H^\prime} = 750$ GeV and $\sqrt{s} = 13$ TeV} we get
\begin{eqnarray}
&& \sigma_{pp \to {H^\prime}+X} \cdot Br({H^\prime} \to \gamma \gamma) \approx  \Big(0.03 |c^{H^\prime}_g|^2 \nonumber\\&&+ 0.3 \zeta_{bb{t}{t}} |c^{H^\prime}_t|^2\Big) \times  \frac{|c_\gamma^{H^\prime}|^2 }{|c^{H^\prime}_t|^2}\text{ fb}.
 \label{CROSS}
\end{eqnarray}
We assume, that $\zeta_{bb{t}{t}}\le 0.1$ while our numerical estimates give $|c^{H^\prime}_t|^2\sim 1$, $|c^{H^\prime}_g|^2\sim 1$, $|c^{H^\prime}_\gamma|^2 < 1$. \revisionA{For all considered choices of parameters with different values of $M_{H^\prime}$ at $\sqrt{s} = 13$ TeV we present the values of the cross section $\sigma_{pp \to {H^\prime}+X \to \gamma \gamma + X} = \sigma_{pp \to {H^\prime}+X} \cdot Br({H^\prime} \to \gamma \gamma) $ in the captions to Tables \ref{table1}, \ref{table2}, \ref{table4}, \ref{table5}, \ref{table6}, \ref{table7}. We can see, that this cross section is decreased fast when the value of $M_{H^\prime}$ is increased. Overall, we come to conclusion, that at the considered choices of parameters this model gives the values of the cross - section
\begin{equation}
 \sigma_{pp \to {H^\prime}+X} \cdot Br({H^\prime} \to \gamma \gamma) \le 0.03 \, \text{ fb}.
 \label{UB}
\end{equation}}
In our model in addition to the second neutral CP - even Higgs boson there are the other two scalar bosons $H^{\prime\prime}$ and $H^{\prime\prime\prime}$ with the masses, that are of the same order of magnitude, but larger than $M_{H^\prime}$. Our rough estimates of the upper bound on the cross - sections of the creation of those states give the same inequality of Eq. (\ref{UB}).

The present experimental upper bounds on the cross - section at $\sqrt{s} = 13$ TeV are presented in Fig. 6 of \cite{Khachatryan:2016yec}. They give, for example, for the resonance with the width $\sim 0.06 M_{H^{\prime}}$ and the mass $M_{H^{\prime}} \sim 1$ TeV:
\begin{equation}
 \sigma_{pp \to {H^\prime}+X} \cdot Br({H^\prime} \to \gamma \gamma) \le 1\, \text{ fb}.
 \label{pp->SX*branching-value0}
\end{equation}
This value is decreased with the increase of $M_{H^{\prime}}$, and gives the value around $0.1$ fb at $M_{H^{\prime}} \sim 3$ TeV. If the cross - section of the second Higgs approaches this value, then the model of \cite{VZ2015} should be extended. There are in general two possibilities. The first one is to modify the model in such a way that the production cross - section of $H^\prime$ is enhanced. The other way is to provide sufficiently larger value of the partial decay width $\Gamma_{H^\prime \to \gamma \gamma}$.  The first may, in principle be achieved if we assume, that the heavy quarks are created out of the light quarks at the $pp$ collisions.  The second may be achieved following the idea of \cite{F16} that the decay of $H^\prime$ to two photons may go through the virtual loop of composite charged bosons.

\begin{widetext}
\begin{table}
\begin{center}
\begin{small}
\begin{tabular}{|c|c|c|c|c|c|c|c|c|c|c|c|c|c|c|c|c|}
\hline
$\Lambda$ & $g^{1/2}_t$ & $g^{1/2}_{\chi}$ & $g^{1/2}_{t\chi}$ & $m_t$ & $m_\chi$ & $M_H$ & $\eta$  \\
\hline
$2757$ & $260$  & $38$  & $100$ &$174$  & $1800$ & $125$ & $246$ \\
\hline
 $M_{H^\prime} $& $\Gamma_{H^\prime}$ & $M_{H^{\prime\prime}}$ & $\Gamma_{H^{\prime\prime}}$ & $M_{H^{\prime \prime \prime}}$ & $\Gamma_{H^{\prime\prime\prime}}$ & $M_{H^{\pm}}$ & $\Gamma_{H^{\pm}}$ \\
\hline
 $751$ & $236$ & $3601$ & $0.28$  & $2352$ & $378$ &$829$ & $378$   \\
\hline
$C^{H^\prime}_{t}$ & $C^{H^\prime}_{\chi}$  & $C^{H^\prime}_{g}$  & $C^{H^\prime}_{\gamma}$ & $C^{H^\prime}_{W}$ & $C^{H^\prime}_{Z}$ & $C^{H^\prime}_{t_L\chi_R}$ & $C^{H^\prime}_{\chi_L t_R}$ \\
\hline
$\begin{array}{c} 2.853 \\  -0.765 \,i \end{array}$ &  $\begin{array}{c} 0.0641\\-0.0146\, i \end{array}$ & $\begin{array}{c} 2.142\\+2.84\, i \end{array} $ & $\begin{array}{c} 0.480\\ +0.660\, i \end{array}$ &$\begin{array}{c} -0.0863\\-0.0513\, i \end{array}$ & $\begin{array}{c} -0.0863\\-0.0513\, i\end{array}$ & $\begin{array}{c} 1.74 \\+0.496\, i \end{array} $ & $\begin{array}{c}  0.246 \\-0.154\, i\end{array}$   \\
\hline
$C^{H}_{t}$ & $C^{H}_{\chi}$  & $C^{H}_{g}$  & $C^{H}_{\gamma}$ & $C^{H}_{W}$ & $C^{H}_{Z}$ & $C^{H}_{t_L\chi_R}$ & $C^{H}_{\chi_L t_R}$ \\
\hline
 $0.854$ &  $-0.00442$ & $0.881$ &$-0.577$ & $0.740$ & $0.740$ & $-5.16 $ & $0.469$   \\
\hline
$C^{H,SM}_{t}$ & $C^{H,SM}_{\chi}$  & $C^{H,SM}_{g}$  & $C^{H,SM}_{\gamma}$ & $C^{H,SM}_{W}$ & $C^{H,SM}_{Z}$ & $C^{H,SM}_{t_L\chi_R}$ & $C^{H,SM}_{\chi_L t_R}$ \\
\hline
 $1$ &  $0$ & $1.03$ &$-0.814$ & $1$ & $1$ & $0$ & $0$   \\
\hline
\end{tabular}
\end{small}
\caption{In this table we represent the example choice of the parameters of the model that corresponds to $m_\chi= 1800$ GeV, $M_{H^\prime} = 750$ GeV, and $\Gamma_{H^\prime}\approx  240$ GeV.  We represent here parameters $g_{t}, g_\chi, g_{t\chi}, \Lambda$, and the corresponding  values of the mass of the second Higgs $M_{H^\prime}$, the mass of the third Higgs $M_{H^{\prime\prime}}$, and the mass of the charged scalar boson $H^{\pm}$. We also represent here the decay widths to the pairs consisted of $\chi, t, b$. All dimensional values are given in GeV. Besides, we represent here the effective couplings of the first and the second Higgs bosons to the fields of the Standard Model and to $\chi$ (those couplings are dimensionless). We also represent here for the comparison the decay constants of the SM Higgs boson (the corresponding values are supplemented by the superscript $SM$). \revisionA{For this choice of parameters $\sigma_{pp \to {H^\prime}+X \to \gamma \gamma + X} \approx 0.032$ fb while $\sigma_{pp \to {H^\prime}+X } \approx 3.1$ pb at $\sqrt{s} = 13$ TeV.}} \label{table1}
\end{center}
\end{table}
\end{widetext}

\subsection{The example choices of parameters that provide $M_H = 125$ GeV, $M_{H^\prime}\approx 750$ GeV, $1200$ GeV, $1600$ GeV,  $2000$ GeV, and  $\Gamma_{H^\prime} \sim 0.3 M_{H^\prime}$}

In this section we represent the example choices of the parameters of the model, which allow to identify the experimentally observed scalar boson with mass $125$ GeV with the CP even goldstone boson of our model, while the second CP even scalar boson mass is around $750$  GeV, $1200$ GeV, $1600$ GeV, and $2000$ GeV with the width $\Gamma_{H^\prime} \approx 0.3 M_{H^\prime}$.

\begin{widetext}
\begin{table}
\begin{center}
\begin{small}
\begin{tabular}{|c|c|c|c|c|c|c|c|c|c|c|c|c|c|c|c|c|}
\hline
$\Lambda$ & $g^{1/2}_t$ & $g^{1/2}_{\chi}$ & $g^{1/2}_{t\chi}$ & $m_t$ & $m_\chi$ & $M_H$ & $\eta$  \\
\hline
$4694$ & $285$  & $62$  & $132$ &$174$  & $1200$ & $125$ & $246$ \\

\hline
 $M_{H^\prime} $& $\Gamma_{H^\prime}$ & $M_{H^{\prime\prime}}$ & $\Gamma_{H^{\prime\prime}}$ & $M_{H^{\prime \prime \prime}}$ & $\Gamma_{H^{\prime\prime\prime}}$ & $M_{H^{\pm}}$ & $\Gamma_{H^{\pm}}$ \\
\hline
 $750$ & $ 198$ & $2403$ & $1.16$  & $1688$ & $238$ &$841$ & $326$   \\
\hline
$C^{H^\prime}_{t}$ & $C^{H^\prime}_{\chi}$  & $C^{H^\prime}_{g}$  & $C^{H^\prime}_{\gamma}$ & $C^{H^\prime}_{W}$ & $C^{H^\prime}_{Z}$ & $C^{H^\prime}_{t_L\chi_R}$ & $C^{H^\prime}_{\chi_L t_R}$ \\
\hline
$\begin{array}{c} 2.60 \\  -0.561 \,i \end{array}$ &  $\begin{array}{c} 0.0768\\-0.0147\, i \end{array}$ & $\begin{array}{c} 1.81\\+2.65\, i \end{array} $ & $\begin{array}{c} 0.413\\ +0.614\, i \end{array}$ &$\begin{array}{c} -0.0918\\-0.0243\, i \end{array}$ & $\begin{array}{c} -0.0918\\-0.0243\, i\end{array}$ & $\begin{array}{c} 1.32 \\+0.192\, i \end{array} $ & $\begin{array}{c}  0.346\\-0.159\, i\end{array}$   \\
\hline
$C^{H}_{t}$ & $C^{H}_{\chi}$  & $C^{H}_{g}$  & $C^{H}_{\gamma}$ & $C^{H}_{W}$ & $C^{H}_{Z}$ & $C^{H}_{t_L\chi_R}$ & $C^{H}_{\chi_L t_R}$ \\
\hline
 $0.896$ &  $-0.00584$ & $0.924$ &$-0.595$ & $0.767$ & $0.767$ & $-3.57 $ & $0.485$   \\
\hline
\end{tabular}
\end{small}
\caption{In this table we represent the example choice of the parameters of the model that corresponds to $m_\chi= 1200$ GeV, $M_{H^\prime} = 750$ GeV,  and $\Gamma_{H^\prime}\approx 200$ GeV. All dimensional values are given in GeV.  \revisionA{For this choice of parameters $\sigma_{pp \to {H^\prime}+X \to \gamma \gamma + X} \approx 0.025$ fb  while $\sigma_{pp \to {H^\prime}+X } \approx 2.5$ pb at $\sqrt{s} = 13$ TeV. }} \label{table2}
\end{center}
\end{table}
\end{widetext}

First of all, let us list the relations to be provided by our choice of parameters:
\begin{enumerate}

\item{$\eta \approx 246$ GeV}

\item{$M_H \approx 125$ GeV}

\item{${\rm Re}\, M_{H^{\prime}} \approx 750 {\rm GeV} $, $1200$ GeV, $1600$ GeV, and $2000$ GeV}

\item{$\Gamma_{H^\prime} \sim 0.3 \, M_{H^\prime}$}

\end{enumerate}

We have in total $5$ parameters to be fixed:  $\Lambda$, $m_\chi$,
$g_\chi$, $g_t$, $g_{t\chi}$,  and $4$ equations to be solved.
In order to fix the initial parameters of the model corresponding to the chosen values of $m_\chi$ and $\Gamma_{H^\prime}$ (dominated by the partial decay width for $H^\prime \to \bar{t}t$) we use the numerical methods based on the gradient descend algorithm. Those numerical methods allow us to fix the values of ${\rm Re}\, M_{H^\prime} \approx  750 {\rm GeV}$, $1200$  GeV, $1600$ GeV,  $2000$ GeV, and $\eta = 246$ GeV, $M_H = 125$ GeV with the accuracy of about $1$ per cent.

We considered the ranges of the values of $m_\chi$:
$$ 600 \, {\rm GeV} \le m_\chi \le 6000\, {\rm GeV}$$
and the ranges of $\Gamma_{H^\prime}$:
$$ 20\,{\rm GeV} \le \Gamma_{H^\prime} \le 700\, {\rm GeV}$$
For each value of $m_\chi$ and $\Gamma_{H^\prime}$ there exists the discrete sequence of the values of the cutoff $\Lambda$ entering our loop integrals, for which the values $M_H = 125$ GeV, $\eta = 246$ GeV, ${\rm Re}\, M_{H^\prime} =  750 {\rm GeV}$,  $1200$ GeV, $1600$ GeV, and $2000$ GeV are provided (there exists the corresponding choice of $g_\chi$, $g_t$, $g_{t\chi}$). Typically in those  sequences the values of $\Lambda$ differ by about the order of magnitude starting from about $\sim 3$ TeV.

It appears, that the necessity to reproduce the observed coupling constants of the $125$ GeV Higgs boson \cite{Khachatryan:2016vau} constraints essentially the admitted values of the parameters of our model. In particular, only the solutions with sufficiently large values of $\Gamma_{H^\prime}$ are relevant. Besides, we observe, that as it was predicted in \cite{VZ2015} the values of the decay constants of the $125$ GeV Higgs boson become closer to the SM values as the ratio $m_t/m_\chi$ is decreased.

We represent in Tables \ref{table1},\ref{table2}, \ref{table4}, \ref{table5}, \ref{table6}, \ref{table7} the example choices of the parameters of our model that illustrate the dependence of various observed quantities on $m_\chi$, $M_{H^\prime}$, and $\Gamma_{H^\prime}$.
It is worth mentioning, that parameters $g_{t},g_\chi, g_{t\chi}$ represented here are not the bare parameters of the model. They are related via the rotation with the angle $\alpha$ to the bare parameters   $g^{(0)}_{t},g^{(0)}_\chi, g^{(0)}_{t\chi}$ (for the details see \cite{VZ2015}).

\begin{widetext}
\begin{table}
\begin{center}
\begin{small}
\begin{tabular}{|c|c|c|c|c|c|c|c|c|c|c|c|c|c|c|c|c|}
\hline
$\Lambda$ & $g^{1/2}_t$ & $g^{1/2}_{\chi}$ & $g^{1/2}_{t\chi}$ & $m_t$ & $m_\chi$ & $M_H$ & $\eta$  \\
\hline
$11286$ & $273$  & $80.1$  & $144.6$ &$174$  & $1600$ & $125$ & $246$ \\

\hline
 $M_{H^\prime} $& $\Gamma_{H^\prime}$ & $M_{H^{\prime\prime}}$ & $\Gamma_{H^{\prime\prime}}$ & $M_{H^{\prime \prime \prime}}$ & $\Gamma_{H^{\prime\prime\prime}}$ & $M_{H^{\pm}}$ & $\Gamma_{H^{\pm}}$ \\
\hline
 $750$ & $ 160$ & $3201$ & $0.12$  & $2163$ & $196$ &$712$ & $93.7$   \\
\hline
$C^{H^\prime}_{t}$ & $C^{H^\prime}_{\chi}$  & $C^{H^\prime}_{g}$  & $C^{H^\prime}_{\gamma}$ & $C^{H^\prime}_{W}$ & $C^{H^\prime}_{Z}$ & $C^{H^\prime}_{t_L\chi_R}$ & $C^{H^\prime}_{\chi_L t_R}$ \\
\hline
$\begin{array}{c} 2.3364 \\  -0.411 \,i \end{array}$ &  $\begin{array}{c} 0.0246\\-0.003\, i \end{array}$ & $\begin{array}{c} 1.5113\\+2.42\, i \end{array} $ & $\begin{array}{c}  0.337 \\+0.5399\, i\end{array}$  & $\begin{array}{c} -0.0057\\ -0.0011\, i \end{array}$ &$\begin{array}{c} -0.0057\\-0.0011\, i \end{array}$ & $\begin{array}{c} 1.176\\+0.14\, i\end{array}$ & $\begin{array}{c} 0.00284 \\-0.0385\, i \end{array} $   \\
\hline
$C^{H}_{t}$ & $C^{H}_{\chi}$  & $C^{H}_{g}$  & $C^{H}_{\gamma}$ & $C^{H}_{W}$ & $C^{H}_{Z}$ & $C^{H}_{t_L\chi_R}$ & $C^{H}_{\chi_L t_R}$ \\
\hline
 $0.84$ &  $-0.002$ & $0.87$ &$-0.754$ & $0.908$ & $0.908$ & $-3.08 $ & $0.333$   \\
\hline
\end{tabular}
\end{small}
\caption{In this table we represent the example choice of the parameters of the model that corresponds to $m_\chi= 1600$ GeV, $M_{H^\prime} = 750$ GeV,  and $\Gamma_{H^\prime}\approx 160$ GeV. All dimensional values are given in GeV.  \revisionA{For this choice of parameters $\sigma_{pp \to {H^\prime}+X \to \gamma \gamma + X} \approx 0.018$ fb  while $\sigma_{pp \to {H^\prime}+X } \approx 2$ pb at $\sqrt{s} = 13$ TeV. }} \label{table4}
\end{center}
\end{table}
\end{widetext}

The parameters of the model fixed above allow to derive expressions for all composite scalar boson masses except for the ones, which values depend on $b_t,b_\chi,b_{t\chi}$.

It appears, that the other CP even scalar boson appears, which is denoted by $H^{\prime \prime}$. Its mass is given approximately by $2 m_\chi$, and it is composed mostly of the pair $\bar{\chi}\chi$. The remaining neutral CP even scalar boson in \cite{VZ2015} was denoted by $M^{(2)}_{h_th_{\chi}}$. Here we denote its mass by $M_{H^{\prime \prime \prime}}$.  The mass of the charged scalar boson $H^\pm$ was denoted in \cite{VZ2015} by $M^{ (2)}_{H^{\pm}_t, H^{\pm}_{\chi}}$.

The masses of the three CP - odd scalar bosons that may exist in this model were denoted in \cite{VZ2015}
by $M^{(2)}_{A_t A_{\chi}}$ and $M^{(1,2)}_{\pi_{\chi}, \pi_t}$. Those masses depend on the parameters $b_t,b_\chi,b_{t\chi}$, which are not fixed here. We expect those masses to be of the order of a few TeV.

\begin{widetext}
\begin{table}
\begin{center}
\begin{small}
\begin{tabular}{|c|c|c|c|c|c|c|c|c|c|c|c|c|c|c|c|c|}
\hline
$\Lambda$ & $g^{1/2}_t$ & $g^{1/2}_{\chi}$ & $g^{1/2}_{t\chi}$ & $m_t$ & $m_\chi$ & $M_H$ & $\eta$  \\
\hline
$13690$ & $744$  & $161$  & $326$ &$174$  & $6000$ & $125.4$ & $246$ \\

\hline
 $M_{H^\prime} $& $\Gamma_{H^\prime}$ & $M_{H^{\prime\prime}}$ & $\Gamma_{H^{\prime\prime}}$ & $M_{H^{\prime \prime \prime}}$ & $\Gamma_{H^{\prime\prime\prime}}$ & $M_{H^{\pm}}$ & $\Gamma_{H^{\pm}}$ \\
\hline
 $2000$ & $ 700$ & $12000.22$ & $0.0188$  & $7671$ & $927$ &$1985$ & $717$   \\
\hline
$C^{H^\prime}_{t}$ & $C^{H^\prime}_{\chi}$  & $C^{H^\prime}_{g}$  & $C^{H^\prime}_{\gamma}$ & $C^{H^\prime}_{W}$ & $C^{H^\prime}_{Z}$ & $C^{H^\prime}_{t_L\chi_R}$ & $C^{H^\prime}_{\chi_L t_R}$ \\
\hline
$\begin{array}{c} 2.561 \\  -0.704 \,i \end{array}$ &  $\begin{array}{c} 0.0084\\-0.0017\, i \end{array}$ & $\begin{array}{c} 0.036\\+0.938\, i \end{array} $ & $\begin{array}{c}  -0.0027 \\+0.2220\, i\end{array}$  & $\begin{array}{c} -0.0374\\ -0.0553\, i \end{array}$ &$\begin{array}{c} -0.0374\\-0.0553\, i \end{array}$ & $\begin{array}{c} 1.706\\+0.505\, i\end{array}$ & $\begin{array}{c} 0.0011 \\-0.0284\, i \end{array} $   \\
\hline
$C^{H}_{t}$ & $C^{H}_{\chi}$  & $C^{H}_{g}$  & $C^{H}_{\gamma}$ & $C^{H}_{W}$ & $C^{H}_{Z}$ & $C^{H}_{t_L\chi_R}$ & $C^{H}_{\chi_L t_R}$ \\
\hline
 $0.974$ &  $-0.0001$ & $1.005$ &$-0.8085$ & $0.988$ & $0.988$ & $-4.09 $ & $0.126$   \\
\hline
\end{tabular}
\end{small}
\caption{In this table we represent the example choice of the parameters of the model that corresponds to $m_\chi= 6000$ GeV, $M_{H^\prime} = 2000$ GeV,  and $\Gamma_{H^\prime}\approx 700$ GeV. All dimensional values are given in GeV. \revisionA{For this choice of parameters $\sigma_{pp \to {H^\prime}+X \to \gamma \gamma + X} \approx 0.0002$ fb while $\sigma_{pp \to {H^\prime}+X } \approx 5.2$ fb at $\sqrt{s} = 13$ TeV.}} \label{table5}
\end{center}
\end{table}
\end{widetext}

One can see, that for the example choices of the parameters of our model presented in Tables  \ref{table4}, \ref{table5}, \ref{table6}, \ref{table7}  the decay constants $c^H_t,c^H_Z,c^H_W,c^H_g,c^H_\gamma$ for the $125$ GeV $H$ boson deviate from the SM values within the error bars presented in \cite{Khachatryan:2016vau}. At the same time Tables \ref{table1}, \ref{table2} demonstrates, that  when the ratio $m_t/m_\chi$ is increased, the deviation becomes stronger.

It is worth mentioning, that the coupling constants for the interaction of $H$ with the pair of $t$ and $\chi$ are not small. This may, possibly, explain the invisible decay width of the $125$ GeV scalar boson. Besides, looking at the presented sets of parameters one can easily find that both $H$ and $H^\prime$ are composed mostly of the pairs $\bar{t}_L t_R$ and $\bar{t}_L\chi_R$.

\begin{widetext}
\begin{table}
\begin{center}
\begin{small}
\begin{tabular}{|c|c|c|c|c|c|c|c|c|c|c|c|c|c|c|c|c|}
\hline
$\Lambda$ & $g^{1/2}_t$ & $g^{1/2}_{\chi}$ & $g^{1/2}_{t\chi}$ & $m_t$ & $m_\chi$ & $M_H$ & $\eta$  \\
\hline
$6263$ & $405$  & $80$  & $176$ &$174$  & $3500$ & $125$ & $246$ \\

\hline
 $M_{H^\prime} $& $\Gamma_{H^\prime}$ & $M_{H^{\prime\prime}}$ & $\Gamma_{H^{\prime\prime}}$ & $M_{H^{\prime \prime \prime}}$ & $\Gamma_{H^{\prime\prime\prime}}$ & $M_{H^{\pm}}$ & $\Gamma_{H^{\pm}}$ \\
\hline
 $1200$ & $ 420$ & $7000.42$ & $0.045$  & $4453$ & $583$ &$1182.49$ & $453$   \\
\hline
$C^{H^\prime}_{t}$ & $C^{H^\prime}_{\chi}$  & $C^{H^\prime}_{g}$  & $C^{H^\prime}_{\gamma}$ & $C^{H^\prime}_{W}$ & $C^{H^\prime}_{Z}$ & $C^{H^\prime}_{t_L\chi_R}$ & $C^{H^\prime}_{\chi_L t_R}$ \\
\hline
$\begin{array}{c} 2.68 \\  -0.75 \,i \end{array}$ &  $\begin{array}{c} 0.0198\\-0.0043\, i \end{array}$ & $\begin{array}{c} 0.4942\\+1.8658\, i \end{array} $ & $\begin{array}{c}  0.111 \\+0.430\, i\end{array}$  & $\begin{array}{c} -0.0429\\ -0.0595\, i \end{array}$ &$\begin{array}{c} -0.0429\\-0.0595\, i \end{array}$ & $\begin{array}{c} 1.70\\+0.57\, i\end{array}$ & $\begin{array}{c} 0.0216 \\-0.0572\, i \end{array} $   \\
\hline
$C^{H}_{t}$ & $C^{H}_{\chi}$  & $C^{H}_{g}$  & $C^{H}_{\gamma}$ & $C^{H}_{W}$ & $C^{H}_{Z}$ & $C^{H}_{t_L\chi_R}$ & $C^{H}_{\chi_L t_R}$ \\
\hline
 $0.932$ &  $-0.0006$ & $0.96$ &$-0.78$ & $0.95$ & $0.95$ & $-4.64 $ & $0.234$   \\
\hline
\end{tabular}
\end{small}
\caption{In this table we represent the example choice of the parameters of the model that corresponds to $m_\chi= 3500$ GeV, $M_{H^\prime} = 1200$ GeV,  and $\Gamma_{H^\prime}\approx 420$ GeV. All dimensional values are given in GeV.  \revisionA{For this choice of parameters $\sigma_{pp \to {H^\prime}+X \to \gamma \gamma + X} \approx 0.001$ fb  while $\sigma_{pp \to {H^\prime}+X } \approx 0.2$ pb at $\sqrt{s} = 13$ TeV.}} \label{table6}
\end{center}
\end{table}
\end{widetext}

For the example choice of parameters corresponding to Table \ref{table1} the production cross - section of the second Higgs boson is given by the sum of Eq. (\ref{productiongg}) and Eq. (\ref{productionbb}). Assuming $\zeta_{ttbb} \ll 0.1$ we neglect the contribution of Eq. (\ref{productionbb}) and  obtain the order of magnitude estimate
\begin{equation}
\sigma^{}_{pp \to {H^\prime}+X} \sim 3.1\, {\rm pb}
\end{equation}
for the production cross section of $H^\prime$ at the second run of LHC with $\sqrt{s}=13$ TeV.  The estimate for the other considered example choices of parameters gives the result presented in the captions to Tables \ref{table1}-\ref{table7}. Roughly, the same quantity at $\sqrt{s} = 8$ TeV is one order of magnitude smaller. The decay of $H^\prime$ is dominated by the channel $t\bar{t}$. This suggests the necessity to search for the new resonance in this channel rather than in the $\gamma \gamma$ channel, where the cross - section is rather weak (about two orders of magnitude smaller than the present experimental constraints). For the choice of parameters represented in Table \ref{table1} (which is, of course, not the only choice) our results indicate that  $\sigma^{}_{pp \to {H^\prime} + X\to t\bar{t} + X} \sim 0.3\, {\rm pb}$ for $\sqrt{s} = 8$ TeV is very close to the experimental upper bound represented in Table 1 of \cite{750REV} and in Fig. 11 of \cite{Aad:2015fna}. This upper bound is decreased when the scalar boson mass is increased, and becomes of the order of $10^{-2}$ pb at $M_{H^\prime} \sim 3$ TeV. Our predictions for the cross - section $\sigma^{}_{pp \to {H^\prime} + X\to t\bar{t} + X}$ for the considered sets of parameters remain smaller than the corresponding upper bound.  In these references the upper bound on the decays to $\bar{t}t$  is given for the first LHC run only. We expect that the run II data will constrain the value of the cross section in this channel much stronger, and will allow either to confirm or to disfavor the scenario presented in the present paper.

\begin{widetext}
\begin{table}
\begin{center}
\begin{small}
\begin{tabular}{|c|c|c|c|c|c|c|c|c|c|c|c|c|c|c|c|c|}
\hline
$\Lambda$ & $g^{1/2}_t$ & $g^{1/2}_{\chi}$ & $g^{1/2}_{t\chi}$ & $m_t$ & $m_\chi$ & $M_H$ & $\eta$  \\
\hline
$7156$ & $537.7$  & $129.5$  & $260.75$ &$174$  & $2500$ & $124.8$ & $246$ \\

\hline
 $M_{H^\prime} $& $\Gamma_{H^\prime}$ & $M_{H^{\prime\prime}}$ & $\Gamma_{H^{\prime\prime}}$ & $M_{H^{\prime \prime \prime}}$ & $\Gamma_{H^{\prime\prime\prime}}$ & $M_{H^{\pm}}$ & $\Gamma_{H^{\pm}}$ \\
\hline
 $1600$ & $ 640$ & $5001.75$ & $0.52$  & $3297$ & $423$ &$1645$ & $720$   \\
\hline
$C^{H^\prime}_{t}$ & $C^{H^\prime}_{\chi}$  & $C^{H^\prime}_{g}$  & $C^{H^\prime}_{\gamma}$ & $C^{H^\prime}_{W}$ & $C^{H^\prime}_{Z}$ & $C^{H^\prime}_{t_L\chi_R}$ & $C^{H^\prime}_{\chi_L t_R}$ \\
\hline
$\begin{array}{c} 2.76 \\  -0.79 \,i \end{array}$ &  $\begin{array}{c} 0.0571\\-0.0157\, i \end{array}$ & $\begin{array}{c} 0.122\\+1.364\, i \end{array} $ & $\begin{array}{c}  0.033 \\+0.317\, i\end{array}$  & $\begin{array}{c} -0.0470\\ -0.0506\, i \end{array}$ &$\begin{array}{c} -0.0470\\-0.0506\, i \end{array}$ & $\begin{array}{c} 1.3887\\+0.3980\, i\end{array}$ & $\begin{array}{c} 0.2422 \\-0.1543\, i \end{array} $   \\
\hline
$C^{H}_{t}$ & $C^{H}_{\chi}$  & $C^{H}_{g}$  & $C^{H}_{\gamma}$ & $C^{H}_{W}$ & $C^{H}_{Z}$ & $C^{H}_{t_L\chi_R}$ & $C^{H}_{\chi_L t_R}$ \\
\hline
 $0.969$ &  $-0.001$ & $1.0002$ &$-0.756$ & $0.938$ & $0.938$ & $-3.837 $ & $0.270$   \\
\hline
\end{tabular}
\end{small}
\caption{In this table we represent the example choice of the parameters of the model that corresponds to $m_\chi= 2500$ GeV, $M_{H^\prime} = 1600$ GeV,  and $\Gamma_{H^\prime}\approx 640$ GeV. All dimensional values are given in GeV. \revisionA{For this choice of parameters $\sigma_{pp \to {H^\prime}+X \to \gamma \gamma + X} \approx 0.00017$ fb while $\sigma_{pp \to {H^\prime}+X } \approx 31$ fb at $\sqrt{s} = 13$ TeV.}} \label{table7}
\end{center}
\end{table}
\end{widetext}

\section{Conclusion and discussions}

\label{sectconclusions}

To conclude, we considered the modified model of top quark condensation proposed in \cite{VZ2015}. To calculate various physical quantities we restrict ourselves by the leading order in the $1/N_c$ expansion, that is in practise the one - loop approximation.  In principle, there exist the higher order corrections to those expressions, but these corrections depend on the way the theory is regularized. There is the commonly accepted methodology for working with the Nambu - Jona - Lasinio (NJL) theories with the non - renormalizable four fermion interaction. According to this methodology only the one - loop results are to be taken into account while the higher order ones are simply disregarded. That means, that the given model is considered rather as the phenomenological model, and not as the true field theory. The extensive discussion of this issue may be found in \cite{VolovikZubkovHiggs,VZ2015}, and in the references therein. Notice, that the example choices of the parameters of our model considered in the present paper reveal the analogy with the NJL model of QCD \cite{NJLQCD} because the ultraviolet cutoff entering our expressions remains of the order of the dynamical mass $m_\chi$. The situation, when the ultraviolet cutoff is not essentially larger than the value of the dynamical fermion mass is often considered as the condition that the NJL model gives a reasonable approximation to the more fundamental theory \cite{cvetic,cvetic_}.

We demonstrate, that the given model is able, in principle, to describe both the $125$ GeV Higgs boson $H$ and the additional more heavier composite scalar boson. For the definiteness we considered the choice of parameters that provides the value $M_{H^\prime} = 750 {\rm GeV} $  (this value corresponds to the excess of events that recently caused the boom of the theoretical papers, but which was not yet confirmed by the latest data). Besides, we consider the example choices of parameters with $M_{H^\prime} \approx 1200$ GeV, $M_{H^\prime} \approx 1600$ GeV, and $M_{H^\prime} \approx 2000$ GeV. We considered several possible values of the mass of the heavy fermion of the order of $1$ TeV, and tune the other parameters of the model in order to achieve the values of the total width around $\sim 0.3 M_{H^\prime}$.

We found that without any modifications the model of \cite{VZ2015} provides the value of the cross section $\sigma_{pp\to H^\prime + X \to \gamma \gamma+X}$, which is \revisionA{essentially smaller than the observed upper bound. Moreover, when $M_{H^\prime}$ is increased, this cross section is decreased fast.} 

In the given interval of parameters our model clearly predicts the
extra $CP$ even neutral scalar boson with the mass around $2 m_\chi$ and with very small width (of the order of $1$ GeV and smaller). The remaining CP even neutral scalar has the mass of the order of several TeV. The production cross section for those two states may be estimated in the similar way to our estimate of $\sigma_{pp\to H^\prime + X \to \gamma \gamma+X}$. For the considered values of parameters it is also essentially smaller, than the present experimental upper bound. Besides, we observe, that the charged scalar boson has mass around $M_{H^\prime}$.   The masses of the three CP - odd neutral scalars (i.e. the neutral pseudo -scalars) that might exist in the model, depend on the extra three parameters of the theory and are, therefore, not constrained by the values of $M_{H^\prime}$ and $\Gamma_{H^\prime}$.

The decay of $H^\prime$ in our model is dominated by the $t\bar{t}$ channel. The value of the cross - section $\sigma^{}_{pp \to {H^\prime} + X\to t\bar{t} + X}$  appears to be rather large. For example, for the particular choice of parameters represented in Table \ref{table1} our results indicate that $\sigma^{}_{pp \to {H^\prime} + X\to t\bar{t} + X} \sim 0.3\, {\rm pb}$ for $\sqrt{s} = 8$ TeV, which is close to the experimental upper bound  represented in Table 1 of \cite{750REV}. At $\sqrt{s} = 13$ TeV we predict $\sigma^{}_{pp \to {H^\prime} + X\to t\bar{t} + X} \sim 3.1 {\rm pb}$ for the same choice of parameters. The similar situation takes place for the other considered choices of parameters mentioned in  Tables \ref{table2}, \ref{table4}, \ref{table5}, \ref{table6}, \ref{table7}. This indicates, that the run II data will allow either to confirm of to disfavor the scenario presented in the present paper basing on the analysis of the $t\bar{t}$ channel.


One of us (Z.V.K.) would like to thank A.Yu. Kotov for the very useful discussion of numerical methods. M.A.Z. kindly acknowledges useful discussions and private communications with G.E.Volovik, V.A.Miransky, V.I.Zakharov, M.N.Chernodub, S.Nikolis, and  the authors of \cite{F16}. The work of Z.V.K. was supported by Russian Science Foundation Grant No 16-12-10059.
The part of the work of  M.A.Z. performed in Russia  was supported by Russian Science Foundation Grant No 16-12-10059 (Sections  \ref{seesaw}, \ref{SectPhenomenology}) while the part of the work made in France (section \ref{SectPhen}) was supported by Le Studium Institute of Advanced Studies.

\end{document}